\documentclass[final,leqno,onefignum,onetabnum]{article}

\usepackage{amssymb,amsfonts,amsmath,graphicx}
\usepackage[a4paper, total={6in, 8in}]{geometry}
\DeclareGraphicsExtensions{.pdf,.png,.jpg}

\title{Smoluchowski reaction kinetics for reactions of any order} 

\author{Mark B. Flegg}
\date{Monash University, Clayton, Victoria, Australia
{mark.flegg@monash.edu}. }

\begin{document}
\maketitle


\begin{abstract}
In 1917, Marian von Smoluchowski presented a simple mathematical description of diffusion-controlled reactions on the scale of individual molecules. His model postulated that a reaction would occur when two reactants were sufficiently close and, more specifically, presented a succinct relationship between the relative proximity  of two reactants at the moment of reaction and the macroscopic reaction rate. Over the last century, Smoluchowski reaction theory has been applied widely in the physical, chemical, environmental and, more recently, the biological sciences. Despite the widespread utility of the Smoluchowski theory, it only describes the rates of second order reactions and is inadequate for the description of higher order reactions for which there is no equivalent method for theoretical investigation. In this paper, we derive a generalised Smoluchowski framework in which we define what should be meant by proximity in this context when more than two reactants are involved. We derive the relationship between the macroscopic reaction rate and the critical proximity at which a reaction occurs for higher order reactions. Using this theoretical framework and using numerical experiments we explore various peculiar properties of multimolecular diffusion-controlled reactions which, due to there being no other numerical method of this nature, have not been previous reported.
\end{abstract}



\pagestyle{myheadings}
\thispagestyle{plain}
\markboth{Generalised Smoluchowski reaction kinetics}{Generalised Smoluchowski reaction kinetics}

\section{Introduction}

In late 1916, a year before his death, Marian von Smoluchowski, one of the pioneers of statistical physics, submitted a paper titled \textit{Versuch einer mathematischen Theorie der Koagulationskinetik kolloider L\"{o}sungen} (An attempt for a mathematical theory of coagulation kinetics of colloidal solutions) to Zeitschrift f\"{u}r Physikalische Chemie \cite{Smol1917}. In Smoluchowski's paper he imagined a large dilute system of hard spherical particles moving independently with Brownian motion. He wondered what would happen if particles could stick together whenever they `touch'.

\begin{figure}[!ht]
\begin{center}
\includegraphics[width=0.5\columnwidth]{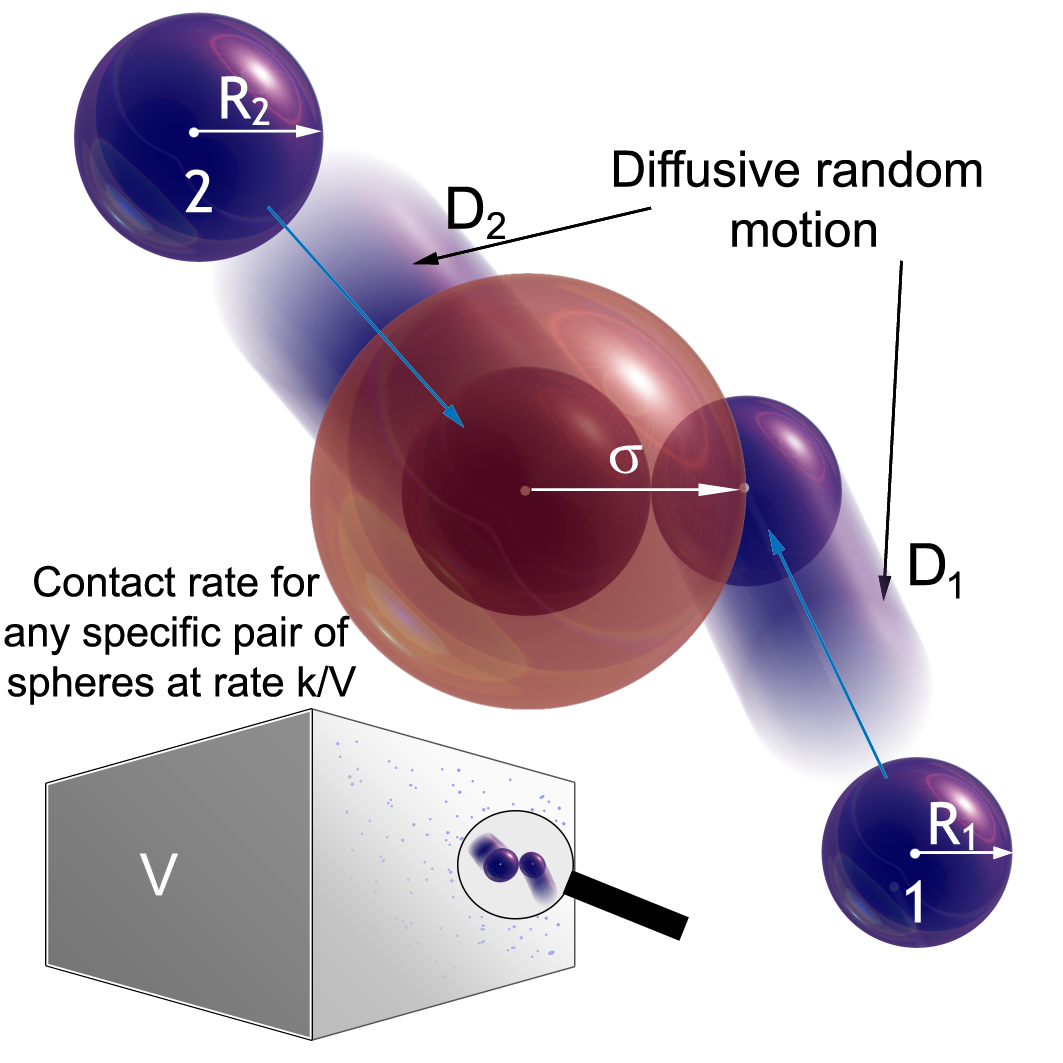}
\caption{Smoluchowski model of diffusing spheres coming into contact. Within a large volume $V$ containing a dilute, well-mixed population of diffusing spheres, contact rates per unit time between any two spheres, 1 and 2, will be dependent on the diffusion constants and radii of each of the spheres and inversely proportional to the volume. The relationship between these parameters and the rate of contact is given by Equation (\ref{smol}).}\label{SmolFig}
\end{center}
\end{figure}

Using the simple example of a closed system containing two diffusing spheres with radii denoted $R_1$ and $R_2$ respectively we can state the famous Smoluchowski result rather succinctly: Diffusing hard spheres placed randomly in a sufficiently large container of volume $V$ will (after an initial period) come into contact at a constant rate $K$ per unit time given by 
\begin{equation}\label{smol}
K= \frac{k}{V} = \frac{4\pi(D_1+D_2)(R_1+R_2)}{V}= \frac{4\pi \hat{D} \sigma }{V},
\end{equation}
where $D_1$ and $D_2$ denote the respective Einstein diffusion coefficients of spheres $1$ and $2$. It is not uncommon to see this result stated independently of the volume, $k = 4\pi \hat{D} \sigma$, and in terms of the relative diffusion coefficient $\hat{D} = D_1+D_2$ and/or the joint effective radius (the distance between the sphere centres at the moment of contact) $\sigma = R_1+R_2$. The Smoluchowski model of diffusing hard spheres is presented diagrammatically in Fig. \ref{SmolFig}.

Smoluchowski's discovery had a very obvious and immediate impact for coagulation theory and is a fundamental theoretical result used in fields such as aerosol physics \cite{Hidy2013}.

As is the case for many simply stated mathematical relations, his equation is more profound than its initial application and has indeed been applied broadly in the last hundred years. Indeed, Smoluchowski himself postulated that the relationship that he had discovered may help in the understanding of chemical reactions \cite{Smol1917}. This is not immediately apparent. Whilst large molecules seemingly move according to Brownian motion in stationary fluids, most cannot adequately be described as hard spheres. Furthermore, chemical reactions do not occur purely as a result of random collisions but rather complex molecular interactions which often require careful analysis and/or numerical simulation. Theorising that a reaction rate $k$ for a chemical reaction can be determined using only knowledge of the molecular kinetics $\hat{D}$ and relative molecular sizes $\sigma$ at first glance seems contrary to the standard model for chemical reactions and molecular interactions. At the very least, the definition of $\sigma$ as a physical size of molecules needs to be relaxed if one is to extend Smoluchowski's theory to include chemical reactions.

There has been a number of attempts to reproduce the result of Smoluchowski throughout the 20th century where the hard spheres are replaced with molecules and interact with other molecules in a fluid. Notably, the introduction of additional physical mechanisms such as intermolecular forces by Kramers \cite{Kramers1940} and Debye \cite{Debye1942} and, later, hydrodynamic effects \cite{Honig1971, Wolynes1976} into the problem resulted in a rate $k$ that looks almost identical to that derived earlier by Smoluchowski with only an alteration to the definition of $\sigma$. The parameter $\sigma$ needed to be redefined and contained information about the nature of the intermolecular forces and/or the hydrodynamic effect considered, but nonetheless was a constant parameter for a given reaction. Whilst historically $\sigma$ is referred to as the Smoluchowski reaction radius, it should more physically be described simply as a reaction parameter. Physically interpreting $\sigma$ merely as a parameter, which summarises the molecular interaction, results in the extension of the Smoluchowski relation to diffusion-controlled reactions. Smoluchowski kinetics has henceforth become an important classical result in chemical reaction theory \cite{Hanggi1990,Turner2004}. Whilst $\sigma$ does not represent a physical contact distance in chemical applications, using the analogous interpretation of Smoluchowski, a complex chemical system made up of many molecules of diffusing reactant species can be reduced, using simulation, to a system of point particles which react when the centre of masses get within a corresponding proximity $\sigma$ of each other (the red sphere in Fig. 1). The value of $\sigma$ used in the simulation can be calculated from the known macroscopic reaction rate $k$ using Equation (\ref{smol}) and often differs wildly from actual molecular sizes.

One particular area of research with which the Smoluchowski relation has had significant impact in the last few decades is biology. This is unsurprising as a vast number of biomolecular systems involve dilute and minute diffusing molecular populations undergoing continuous reaction. Understanding how these biological systems operate is complicated and is in itself a whole field of research; systems biology. The Smoluchowski result has provided a very powerful tool for theoretical investigation of microscopic biochemical reaction-diffusion processes.

One of the greatest limitations to using Smoluchowski reaction kinetics, particularly in the biological sciences, is the fact that the theory only considers the interaction of two particles/molecules at a time. In biology, for example, proteins form multimolecular complexes with very specific biological function (for example, ligand-receptor complexes, transcriptional complexes and other more specific complexes such as the $\beta$-catenin destruction complex). Of course, these multimolecular reactions are usually not true high order reactions. Instead, they are a result of multiple bimolecular reactions. Often, a reaction-diffusion system is modelled using a system of bimolecular and unimolecular reactions and a long list of chemical species and intermediate complexes until it becomes apparent that the number of chemical species may be reduced by identifying high order reactions as a result of fast and slow kinetics. For example, in their seminal paper, Lee \textit{et al}. present an experimentally validated model for the Wnt signalling pathway using an ODE system of 15 chemical species and many bimolecular interactions \cite{lee}. Subsequent analysis of this system by Kruger and Heinrich identified differences in timescales associated with chemical processes involved in the Lee model and by appropriately introducing higher order reactions reduced the number of chemical species from 15 to 7 \cite{Kruger}. Deterministic models of intracellular networks like that of Lee \cite{lee} and Kruger and Heinrich \cite{Kruger} are extremely common in mathematical biology and often omit noise which may be added extrinsically in the form of a stochastic differential equation (SDE) or intrinsically in the form of an agent based simulation. Agent based simulations have the additional advantage over SDEs since they also naturally include spatio-temporal phenomena such as transport-derived time delays. As in the case of the Wnt signalling pathway, Tymchyshyn and Kwiatkowska \cite{Tymchyshyn} and Wawra \textit{et al}. \cite{wawra} respectively show that noise and time delays in the Wnt signalling process can be responsible for substantial oscillations in $\beta$-catenin. Furusawa and Kaneko suggest that oscillations such as these may inform differentiation decision making in cells and quantitative models that capture these effects are critical to our theoretical understanding of cellular processes \cite{furusawa}. Furthermore, protein-protein interactions in the cell are largely transport-limited reactions and thus often require a reaction-diffusion model. There is a clear need, therefore, for agent-based high order reaction-diffusion simulation algorithms. There are two main types of agent-based simulation techniques for reaction-diffusion systems. The first is a lattice-based technique, an extension of the 1977 algorithm by Daniel Gillespie \cite{danG}, known as the spatial Gillespie algorithm. Famously, the spatial Gillespie algorithm does not converge as the lattice is made fine and other more complicated techniques are required \cite{CRDME}. The second is a simulation technique that models molecules as point particles in a continuous domain and reactions occur due to rules in a Smoluchowski framework \cite{Andrews2004}, but these types of simulation are currently limited, as previously mentioned, by the lack of utility for multimolecular reactions. Since the Smoluchowski framework is currently only applicable to diffusive systems with bimolecular or unimolecular reactions they have only found significant biological application in a select few processes. The Min system \cite{ulrich} and the MAPK pathway \cite{eGFRD} are two common examples from the literature.

        To understand why the classical Smoluchowski model is inadequate to describe the multiple bimolecular reactions leading to complex formation on the scale of individual molecules in examples like that of the Wnt signalling pathway and other intracellular networks, consider an example reaction between molecules $A$, $B$ and $C$. Hypothetically, consider a reaction first between $A$ and $B$, creating a short-lived chemical complex with a very high affinity for $C$. What does Smoluchowski have to say about this process? There are two reactions, between $A$ and $B$ and then between the $AB$ complex and $C$. For the first of these reactions, Smoluchowski theory is adequate. However, the `bimolecular reaction' between $AB$ and $C$ occurs at a very fast rate. The effective Smoluchowski radius that needs to be defined for such a rate to occur, according to Equation (\ref{smol}), is very large. As a result, reactant candidate molecules of $C$ may not even be physically in the vicinity of the $AB$ complex and yet, according to the model, they are forced to somehow, instantaneously, react. Thus, because one chemical reaction occurs on a short timescale, the assumptions that are required for diffusion-limited Smoluchowski theory break down, despite the overall high order reaction being diffusion-limited.
        
      Throughout mostly the late 20th century, there was a lot of theoretical interest in multimolecular reactions. There has been extensive study of the Smoluchowski coagulation equation for multimolecular reactions. This equation describes the rate of change of concentration of particles of particular, often discrete, size distributions and considers spatial effects implicitly in the form of a reaction/coagulation kernel \cite{jiangpaper2,jiangpaper3,krapiviskypaper1}. Different coagulation kernels are usually studied separately and represent the rate of reaction as a function of particle sizes (for example, the sum kernel model suggests that $N$-body collisions occur at a rate proportional to the sum of the sizes of all $N$ particles). These kernels are often heuristically based off Smoluchowski's relation for 2-body interactions, the $N$-body version of which has not been studied in detail. The $N$-body reaction has been studied predominantly in modelling using the law of mass action and ordinary differential equations. As the number of molecules/particles is reduced, noise plays an important role in the reaction kinetics and has been studied extrinsically \cite{Kang}. Explicit spatial considerations for $N$-body reaction-diffusion processes have also been studied. Most of these studies focus on homogeneous reactions (reactions with only one type of reactant) and are studied on a lattice \cite{Leepaper4} (often only in one dimension \cite{privmanpaper5,privmanpaper6,krappaper1}). To date, there has been no study of general $N$-body diffusion-limited reactions which is formulated in the same way as Smoluchowski's ground-breaking 1917 paper \cite{Smol1917}.

Motivated by catalytic trimolecular reactions, Oshanin and others in the 1990s attempted to find the Smoluchowski condition for 3-body reactions \cite{Oshanin1995}. However, the analysis was not easily scalable to reactions of higher order or in high dimensions. In this manuscript, I will present a generalised Smoluchowski relation which holds for heterogeneous reactions of any order and is consistent with (reduces to) the original work by Smoluchowski published in 1917 in the case when the order of the reaction is two \cite{Smol1917}. I will demonstrate that this theory is capable of quantifying and simulating the kinetics of simple chemical systems on a scale of individual molecules which have been previously unachievable using three test problems; a simple well-mixed 3D multimolecular reaction leading to exponential decay of one chemical species at a prescribed test rate, a steady state birth-death process utilising a trimolecular reaction and, finally, the simulation of Turing patterns.

\section{An overview of a generalised Smoluchowski framework} \label{framework}

According to the Smoluchowski theory of chemical reaction, $\sigma$ is intimately linked with $k$, the macroscopic rate of reaction. The only other parameters governing the rate of reaction are the diffusion constants $D_i$ of each of the molecules involved in the chemical reaction (where $i$ denotes the specific type of molecule). According to the geometric interpretation of the chemical reaction, $\sigma$ describes the proximity molecules need to be from each other in order for reactions to occur at a rate prescribed by $k$ (see Equation (\ref{smol})). In the discussion which is to proceed, I shall therefore consider all molecules/particles to be point particles described only by their diffusion constants $D_i$ and positions $\mathbf{x}_i$. In agreement with Smoluchowski theory, there is no consideration of molecular crowding in this manuscript since I will be assuming dilute systems only.

\subsection{Separation between multiple diffusing point particles}

Standard mathematical derivation of the bimolecular Smoluchowski equation (Equation (\ref{smol})) involves first describing the motion of particle $2$ in the frame of reference of particle $1$ by placing particle $1$ at the origin of a new set of coordinates. In this new frame of reference, coordinates $\boldsymbol{\eta}_2 =  \mathbf{x}_2 - \mathbf{x}_1$ describe the relative proximity of particle 2 compared to particle 1. Describing particles 1 and 2 using the proximity/separation vector $\boldsymbol{\eta}_2$ instead of their individual positions $\mathbf{x}_2$ and $\mathbf{x}_1$ allows for the condition of reaction to be defined easily using just one vector (that is, when $\|\boldsymbol{\eta}_2\| < \sigma$). Furthermore, it can be shown that the point $\boldsymbol{\eta}_2$ diffuses linearly with a diffusion constant of $\hat{D}_2 = D_2 + D_1$.

In order to describe the proximity of three or more point particles the separation vector $\boldsymbol{\eta}_i$ (the relative separation of the $i$-th molecule from the collection of $i-1$ previous molecules) is defined by
\begin{equation}\label{proxvect}
\boldsymbol{\eta}_i = \mathbf{x}_i - \bar{\mathbf{x}}_{i-1},
\end{equation}
where $\bar{\mathbf{x}}_{i-1}$ is the weighted average position of the first $(i-1)$ particles. For reasons discussed in detail in Section \ref{derivtheory}, we choose $\bar{\mathbf{x}}_{i-1}$ to be given by the \textit{centre of diffusion} of the first $(i-1)$ molecules. Since it can be shown that $\bar{\mathbf{x}}_{i-1}$ exhibits Brownian motion (with a diffusion constant of $\bar{D}_{i-1}$ - see later Equation (\ref{CoD_D})), $\boldsymbol{\eta}_i$ also exhibits Brownian motion with a diffusion constant of
\begin{equation}\label{Dhat}
\hat{D}_i = D_i + \bar{D}_{i-1}.
\end{equation}
These results will be derived in more detail in Section \ref{derivtheory}.

\subsection{Center of diffusion of multiple diffusing point particles}

     Consider two point particles $1$ and $2$ undergoing Brownian motion in a three-dimensional space. Whilst it is well known that the separation of these points $\boldsymbol{\eta}_2$ moves with Brownian motion, a somewhat lesser known fact is that the motion is completely independent of the diffusive motion of the centre of diffusion $\bar{\mathbf{x}}_2$ (see Section \ref{derivtheory} for details). This is an important property and underlies the reason for choosing it to represent the weighted average position of any system of particles in this theoretical framework. The centre of diffusion is calculated in a similar way to the centre of mass but instead of using the masses of a system of point particles it uses the inverse of their diffusion constants $d_i = D_i^{-1}$ as weights. The centre of diffusion $\bar{\mathbf{x}}_i$ of molecules $1, 2, \ldots, i$ at positions $\mathbf{x}_1, \mathbf{x}_2, \ldots \mathbf{x}_i$ respectively diffusing with rate constants $D_1, D_2, \ldots D_i$ is given by the equation
\begin{equation}\label{CoD}
\bar{\mathbf{x}}_i = \frac{\sum_{j=1}^i d_j \mathbf{x}_j  }{\sum_{m=1}^i d_m} = \frac{\sum_{j=1}^i \mathbf{x}_jD_j^{-1} }{\sum_{m=1}^i D_m^{-1}}.
\end{equation}
We note also that the diffusion constant of the centre of diffusion $\bar{\mathbf{x}}_i$ is given by 
\begin{equation}\label{CoD_D}
\bar{D}_i = \frac{1}{\sum_m^i d_m} = \frac{1}{\sum_m^i D_m^{-1}}.
\end{equation}
Equation (\ref{CoD_D}) is derived in Section \ref{derivtheory}.

\subsection{The generalised Smoluchowski theory}

In this subsection, the main result of this manuscript, the generalised Smoluchowski theory, is summarised. The full derivation of the generalised Smoluchowski theory is presented in Section \ref{derivtheory}.

Consider a sufficiently large closed system of volume $V$ containing $N$ independently diffusing point molecules with diffusion constants $D_1, D_2, \ldots D_N$ respectively. The positions of the $N$ molecules are uniformly initialised over the volume and denoted by $\mathbf{x}_1, \mathbf{x}_2, \ldots \mathbf{x}_N$ respectively. After a short period, the molecules will come into `contact' at a constant rate $K=k/V^{N-1}$ per unit time where 
\begin{equation}\label{gensmol}
k = \left[\prod_{i=2}^{N}\hat{D}_i^{3/2}\right] \frac{4\pi^{\alpha+1}}{\Gamma(\alpha)}\left(\frac{\sigma}{\sqrt{\Delta_N}} \right)^{2\alpha},
\end{equation}
where $\Gamma(t) = \int_0^\infty x^{t-1} e^{-x} \ \mathrm{d}x$ is the Gamma function, $\hat{D}_i$ is defined by Equation (\ref{Dhat}), $\alpha = (3N-5)/2$, $\sigma$ is the Smoluchowski radius and the scale parameter $\Delta_N$ is given by
\begin{equation}\label{scalefact}
\Delta_N = \frac{\sum_{i=1}^{N} D_i^{-1}}{\sum_{i>m} (D_i D_m)^{-1}}.
\end{equation}
The summation sign on the denominator of Equation (\ref{scalefact}) is taken for all integer combinations of $i\in[2,N]$ and $m\in[1,N-1]$ such that $i$ is strictly greater than $m$. The rate $k$ is given in units of $\text{concentration}^{-(N-1)}$ (alternatively $\text{length}^{3(N-1)}$) per unit time. Of particular note in (\ref{gensmol}) is that the product is taken from $i=2$ to $i=N$. Contact is defined when the proximity of the molecules becomes less than the Smoluchowski radius ($\mathcal{P}_N \leq \sigma$). The proximity of the $N$ molecules is expressed in terms of the separation vectors $\boldsymbol{\eta}_2, \boldsymbol{\eta}_3, \ldots, \boldsymbol{\eta}_N$ as defined by Equation (\ref{proxvect}).
\begin{equation}\label{proximity}
\mathcal{P}_N^2 = \sum_{i=2}^N \frac{\Delta_N}{\hat{D}_i} \|\boldsymbol{\eta}_i\|^2.
\end{equation}
Whilst it may appear that the proximity of $N$ diffusing molecules as defined by Equation (\ref{proximity}) is dependent on the order in which the molecules are labelled, it can be shown that this is not the case. Specifically,
\begin{equation}\label{proximity2}
\mathcal{P}_N^2 = \frac{\sum_{i>m} (D_i D_m)^{-1} ||\mathbf{x}_i-\mathbf{x}_m||^2}{\sum_{i>m} (D_i D_m)^{-1}}
\end{equation}
is a weighted average of the square separation between all molecule pairs (interested readers are directed to the Appendix for a proof of this result). Whilst Equation (\ref{proximity2}) is an interesting geometrical interpretation of $\mathcal{P}_N$, we shall find it more convenient (both analytically and computationally) to use the form given in Equation (\ref{proximity}).  The proximity $\mathcal{P}_N$ is a measure of distance/separation of the $N$ molecules and thus has units of length (as does the reaction radius $\sigma$).

In the case of classical Smoluchowski theory, $N=2$. Thus, $\alpha = 1/2$, $\Gamma(\alpha)=\sqrt{\pi}$ and $\Delta_2 = \hat{D}_2 = \hat{D} = D_1+D_2$. Equation (\ref{gensmol}) therefore simplifies to Equation (\ref{smol}) and the definition of proximity (Equation (\ref{proximity})) simplifies to the separation of the two reactant molecules $\mathcal{P}_2 = \|\boldsymbol{\eta}_2 \| = \|\mathbf{x}_2 - \mathbf{x}_1 \|$.

\begin{figure}[!ht]
\begin{center}
\includegraphics[width=0.7\columnwidth]{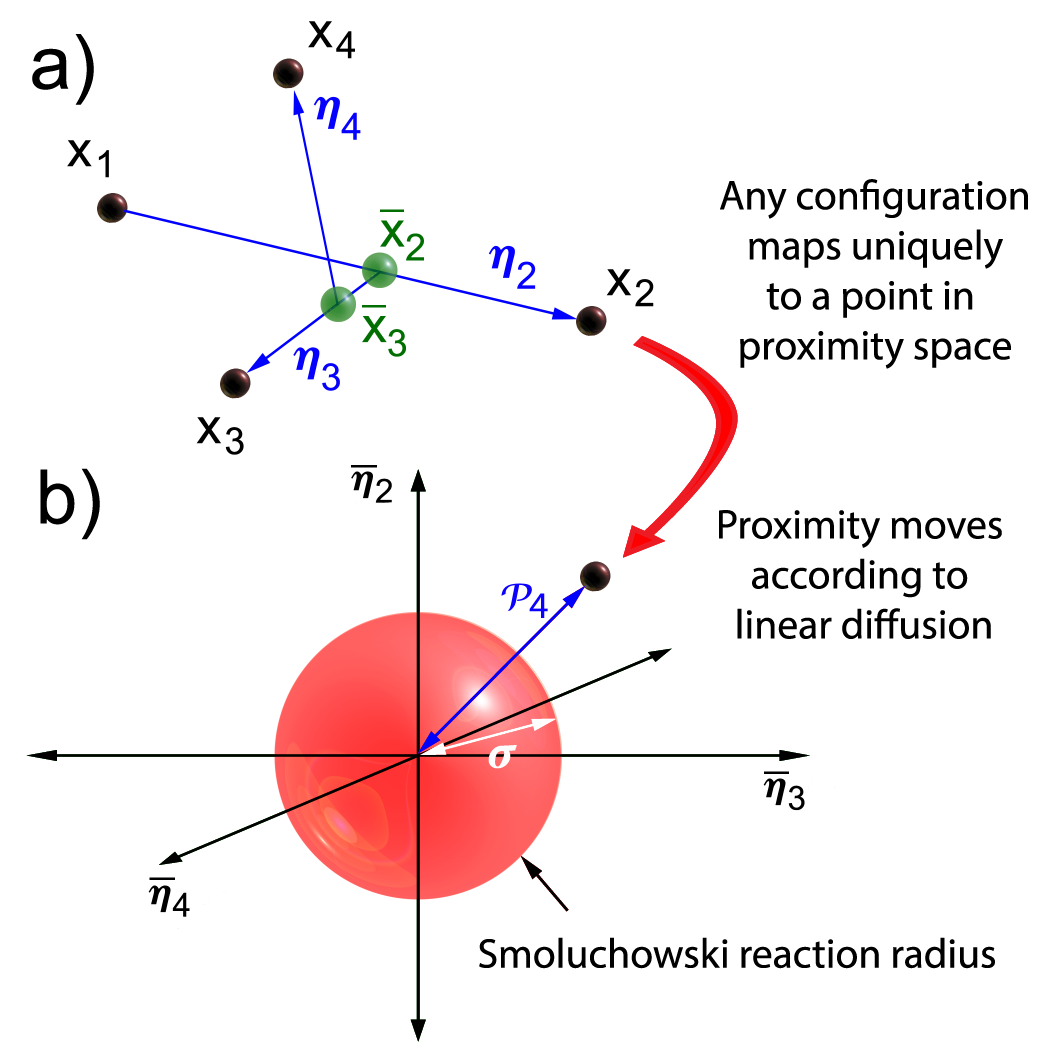}
\caption{According to the generalised Smoluchowski theoretical framework presented in this manuscript, reactions occur when the proximity of the reactants falls within the reaction radius. This reaction radius is diagrammatically represented in this figure using an example of a four-particle system.  a) The proximity is defined by the separation vectors $\eta_i$ shown. These separation vectors use the centres of diffusion (shown in green and calculated using Equation (\ref{CoD})) as reference points. b) Any configuration of reactants in space maps uniquely to a single point/state in proximity space. Note that the proximity space presented in b) is a visualisation only (each of the three axes represent a three coordinate vector and is therefore a visualisation of a 9-dimensional space). In proximity space, the state diffuses according to linear diffusion and a reaction/contact occurs when the proximity falls within the reaction radius $\sigma$ which is related to the reaction rate by Equation (\ref{gensmol}). Note that in this figure, the notation $\bar{\boldsymbol{\eta}}_i = \boldsymbol{\eta}_i\sqrt{\Delta_4/\hat{D}_i}$ is used. }\label{multimolecular}
\end{center}
\end{figure}

A diagrammatic representation of the Smoluchowski reaction radius for a four-particle reaction in proximity space is presented in Fig. \ref{multimolecular}.

\section{Derivation of theoretical results}\label{derivtheory}

In this section, the theoretical results stated in the Section \ref{framework}; an overview of a generalised Smoluchowski framework, will be derived in detail. The key steps will be broken down into subsections so that it is easier to follow.

\subsection{Separation coordinates $\boldsymbol{\eta}$}

We consider here a general $N$-th order reaction occurring between $N$ diffusing molecules. The diffusion constant of the $i$-th molecule is denoted $D_i$. 
The $N$ molecules are initially well-mixed (distributed evenly)
within the arbitrarily large enclosed 3-dimensional domain $\Omega$ of volume
$\iiint_\Omega dv = V$, where $V$ is finite but very large. The volume is enclosed (has Neumann or periodic boundary conditions) so that the reaction becomes inevitable and should occur at a constant exponential rate given by the macroscopic reaction rate. The state of the unreacted system (which is made up of only one possible combination of reactants) at any moment may be described by the set
of 3-dimensional molecule positions $\mathbf{x} = \{\mathbf{x}_1, \mathbf{x}_2, \ldots
, \mathbf{x}_N\}$, where $\mathbf{x}_i\in\Omega$ is the
position vector of the $i$-th molecule. The state space spanned by $\mathbf{x}$ is therefore $[3N]$-dimensional.
 
  Since each of the $N$ molecules diffuse independently, the joint probability density $P$ to find this combination of reactant molecules with
positions $\mathbf{x}_1, \mathbf{x}_2, \ldots, \mathbf{x}_N$ and still in an `unreacted' state is given by
the diffusion equation
\begin{equation}\label{Nparticle2}
\frac{\partial P(\mathbf{x},t)}{\partial t} = \left[\sum_{i=1}^N D_i
\nabla_i^2 \right] P(\mathbf{x},t),
\end{equation}
where $\nabla_i^2$ represents the Laplacian operator with respect to the
coordinates of the position vector $\mathbf{x}_i$.
The condition for an $N$-th order reaction between these $N$ molecules should necessarily be defined in the case when all molecules are sufficiently close to each other (albeit, the metric for `closeness' is not defined at this stage). Since the joint probability density $P$ is zero when the molecules react (are sufficiently close), it is sensible to transform the coordinate system into one in which the coordinates denote the molecules' relative proximity.  We shall perform a linear transformation of the coordinates $\left\{\mathbf{x}_i\right\}_{i=1}^N \rightarrow \left\{\boldsymbol{\eta}_i\right\}_{i=1}^N$,\begin{equation}\label{proxdef}
\boldsymbol{\eta}_i = \sum_{j}^N A_{ij}\mathbf{x}_j.
\end{equation}

 The transformation coefficients $A_{ij}$ are chosen strategically such that  
$\boldsymbol{\eta}_2$ describes the displacement of molecule 2 from molecule 1, $\boldsymbol{\eta}_3$ describes the displacement of molecule 3 from some \textit{weighted average position} (WAP) of molecules 1 and 2, $\boldsymbol{\eta}_4$ describes the displacement of molecule 4 from some WAP of molecules 1, 2 and 3, and so on. Using this new set of coordinates describing the state of the system, the reaction condition will be proposed as some condition in which $\|\boldsymbol{\eta}_2\|, \|\boldsymbol{\eta}_3\|,\ldots , \|\boldsymbol{\eta}_N\|$ are sufficiently small. We shall refer to the state vectors $\boldsymbol{\eta}$ as separation coordinates to differentiate them from the molecule position coordinates $\mathbf{x}$.  It is natural and convenient (for tractability) to chose the transformation coefficients $A_{ij}$ such that the resultant transformed state $\boldsymbol{\eta}$ evolves according to linear diffusion. That is,
\begin{equation}\label{nocross}
 \sum_{i=1}^N D_i  A_{ij}A_{ik} = 0,
\end{equation}
for each $j \neq k$.

Condition (\ref{nocross}) is met, in part, if the WAP is defined by the centre of diffusion. We define the centre of diffusion, $\bar{\mathbf{x}}_i$, of $i$ molecules positioned at $\mathbf{x}_j$ ($j=1,\ldots ,i$) moving with respective diffusion constants $D_j$ using Equation (\ref{CoD})
\begin{equation}\label{CoDA}
\bar{\mathbf{x}}_i = \frac{\sum_{j=1}^i \mathbf{x}_jD_j^{-1} }{\sum_{m=1}^i D_m^{-1}}.
\end{equation}

The vector $\boldsymbol{\eta}_1(\boldsymbol{x})$ remains undefined until now. In order that Condition (\ref{nocross}) holds whilst leaving $\boldsymbol{\eta}_1$ independent of the other vectors $\boldsymbol{\eta}_i$ for $i=2\ldots N$, $\boldsymbol{\eta}_1$ will not denote a displacement between molecules but rather the position of the centre of diffusion for the whole system of $N$ particles.

We therefore have
\begin{eqnarray}\label{trans1}
\boldsymbol{\eta}_1 &=&  \bar{\mathbf{x}}_N \\ \label{trans2}
\boldsymbol{\eta}_i &=& \mathbf{x}_{i} - \bar{\mathbf{x}}_{i-1}, \quad \quad \quad i = 2, 3, \ldots, N.
\end{eqnarray}

The transformation (\ref{proxdef}) denoted by Equations (\ref{trans1}) and (\ref{trans2}) is described by the matrix $\mathbf{A}$ with elements $A_{ij}$ given by
\begin{equation}\label{As}
A_{ij} = \left\lbrace \begin{array}{ccc}
D_j^{-1} \left[\sum_{m=1}^N D_m^{-1} \right]^{-1} & & i=1  \\
-D_j^{-1} \left[\sum_{m=1}^{i-1} D_m^{-1} \right]^{-1} & j< i & i>1 \\
1 & j = i & i>1 \\
0 & j > i & i>1 
\end{array}\right. .
\end{equation}

For this particular linear transformation, the
Jacobian has a determinant of 1 and no scaling of the probability
density is required. Rewriting (\ref{Nparticle2}) in the new coordinate system $\boldsymbol{\eta}$ gives
\begin{equation}\label{Nparticle_independent}
\frac{\partial P_{\boldsymbol{\eta}}(\boldsymbol{\eta},t)}{\partial t} = \left[\sum_{i=1}^N
\hat{D}_i  \hat{\nabla}_i^2  \right] P_{\boldsymbol{\eta}}(\boldsymbol{\eta},t),
\end{equation}
where $P_{\boldsymbol{\eta}}(\boldsymbol{Ax},t) = P(\boldsymbol{x},t)$ and the $\hat{\left.\right.}$ notation denotes the transformed coordinates $\boldsymbol{\eta}$. That is, $\hat{\nabla}_i^2$ is the Laplacian with respect to coordinates of $\boldsymbol{\eta}_i$  and $\hat{D}_i$ is the diffusion constant associated with the state vector $\boldsymbol{\eta}_i$.  
\begin{equation}\label{Dbar}
\hat{D}_i = \sum_{j=1}^N D_j A_{ij}^2 = \left\lbrace \begin{array}{cc}
\bar{D}_N &  i=1  \\
D_i + \bar{D}_{i-1} & i>1  
\end{array}\right. ,
\end{equation}
where $\bar{D}_n$ is the diffusion constant associated with the centre of diffusion of the first $n$ molecules
\begin{equation}\label{realDbar}
\bar{D}_n = \frac{1}{\sum_{i=1}^n D_i^{-1}}.
\end{equation}

\subsection{Reaction radius $\sigma$}

Now that we have defined what is meant by separation coordinates, we will pose a suitable condition which should be placed on the state of the system of $N$ reactants for a reaction to occur.

 According to Equation (\ref{Nparticle_independent}), the state of any unreacted $N$ molecule system, $\boldsymbol{\eta}$, undergoes linear, albeit anisotropic, diffusion. If the $N$ molecules are well mixed throughout the volume $\Omega$, then there is necessarily a uniform probability density for the state $\mathbf{x}$ (and, by extension, $\boldsymbol{\eta}$). In the absence of the perturbation of the joint probability density caused by removal of system states upon reaction, this constant probability density, $P_{\infty}$, can be found by normalisation. That is, $P_{\infty} = 1/V^{N}$.

 Since the condition for reaction should be invariant under translation of the whole system of particles, there is no reaction condition placed on the coordinates of $\boldsymbol{\eta}_1$. That is, the centre of diffusion of the whole system may freely diffuse throughout the volume $\Omega$ without a reaction occurring as a result of such translations. Subsequently, for a well mixed system, the probability density $P_{\boldsymbol{\eta}}$ is independent of $\boldsymbol{\eta}_1$. We therefore integrate Equation (\ref{Nparticle_independent}) over $\Omega$ with respect to the coordinates of $\boldsymbol{\eta}_1$ to reduce the dimensionality of the state space.  
 
 \begin{equation}\label{Nparticle_reduced}
\frac{\partial p(\{\boldsymbol{\eta}\}_2^N,t)}{\partial t} = \left[\sum_{i=2}^{N}
\hat{D}_i  \hat{\nabla}_i^2  \right] p(\{\boldsymbol{\eta}\}_2^N,t),
\end{equation}
where $p(\{\boldsymbol{\eta}\}_2^N,t) = \iiint_\Omega P_{\boldsymbol{\eta}}(\boldsymbol{\eta},t) \ \mathrm{d}v_1 = VP_{\boldsymbol{\eta}}(\{\boldsymbol{\eta}\}_2^N,t)$ (using the notation $\mathrm{d}v_1 = \mathrm{d}\boldsymbol{\eta}_{1;1} \mathrm{d}\boldsymbol{\eta}_{1;2} \mathrm{d}\boldsymbol{\eta}_{1;3}$). Note that the well-mixed constant probability density for the state of the $N$ particle system in this reduced state is given by $p_\infty = 1/V^{N-1}$.

The remaining state vectors $\boldsymbol{\eta}_2,\ldots, \boldsymbol{\eta}_{N}$ represent the relative separations of the $N$ particles. It remains to choose the condition for reaction.

A condition which is placed on the coordinates of the reactants for a reaction to occur is quite arbitrary as it does not reflect specific molecular processes but it merely needs to match a macroscopic reaction rate (the flux of the probability over the reaction boundary must be fixed at a particular macroscopic rate). There are two rules that this boundary must satisfy. The first is that it must describe a condition on the molecules being positioned within some prescribed colocal neighbourhood. The second rule is that the extent of this neighbourhood should be described by one parameter $\sigma$ which is necessarily chosen to match the flux over the boundary with the macroscopic reaction rate. Due to the arbitrariness of this boundary, it is tempting to choose a condition which can be easily described physically. An intuitive physical condition which appears to be a natural extension of the Smoluchowski reaction condition is when  every pair of reactants are within a fixed separation $\sigma$ (that is, the maximum distance between any two molecules is less than $\sigma$).  In their 1995 paper, Oshanin \textit{et al}. impose this very condition to describe three-body reactions \cite{Oshanin1995}. They were successful in matching the free parameter $\sigma$ with the macroscopic reaction rate for trimolecular reactions with the aid of some approximation. The challenge that was faced by the authors was that the chosen condition, although simple and intuitive, did not offer a tractable expression for the flux over the reaction boundary since the distances between diffusing molecules are correlated. The usefulness of the separation coordinates as they have been constructed in this manuscript is that they diffuse independently whilst also describing relative molecular separations (and the separation vectors uniquely have this property, excepting for relabelling permutations). In order to find the flux over the reactive boundary in a tractable form for any number of molecules, we have imposed already that the separation vectors in the state space diffuse linearly, and, since the remaining diffusion is anisotropic (see Equation (\ref{Nparticle_reduced})) we define the reactive boundary to be   
\begin{equation}\label{R}
 \frac{\mathcal{P}_N^2}{\Delta_N} =\sum_{i=2}^{N} \frac{||\boldsymbol{\eta}_i ||^2}{\hat{D}_i} \leq \frac{\sigma^2}{\Delta_N},
\end{equation}
for some \textit{reaction radius} $\sigma$. Defined in Equation (\ref{R}) is the symbol $\mathcal{P}_N$ which we will call the \textit{proximity} of the $N$ molecules and describes the overall separation of the molecules from each other. The scale parameter $\Delta_N$ is completely arbitrary and is placed into the condition so that the proximity and reaction radius are measures of distance. The scale parameter $\Delta_N$ should therefore have units of a diffusion constant. We define $\Delta_N$ using Equation (\ref{scalefact}). Using this definition for $\Delta_N$, the definition of proximity $\mathcal{P}_N$ (and by extension the reaction radius $\sigma$) is invariant of the ordering of the molecule labels. Furthermore, when $N=2$ the Smolouchowski definition of proximity $\mathcal{P}_2 = ||\boldsymbol{\eta}_2 || = \|\mathbf{x}_2-\mathbf{x}_1\|$ is recovered and the Smoluchowski reaction condition (\ref{R}) simplifies to $\|\mathbf{x}_2-\mathbf{x}_1\|\leq \sigma$. Finally, it can be shown that the physical definition of this reaction boundary condition is to ensure that the average square distance between all pairs of molecules, weighted by the inverse product of the molecule pair diffusion constants, is no greater than the square of the reaction radius $\sigma$. A proof of this physical definition for $\mathcal{P}_N$ as well as a justification for using Equation (\ref{scalefact}) to define $\Delta_N$ is presented in the Appendix.
The choice of $\sigma$ is determined by the macroscopic multimolecular reaction rate $k$ measured in expected number of reactions per unit time per unit volume per unit concentration of molecules of each of the $N$ reactants.  We shall see that the relationship between $k$ and $\sigma$ also matches that of Smoluchowski for $N=2$.

\subsection{The relationship between reaction radius $\sigma$ and reaction rate $k$}

 The addition of a reaction condition introduces an absorbing boundary in Equation (\ref{Nparticle_reduced}). We shall define the region of the state space corresponding to the reaction condition (\ref{R}) as $\hat{\Omega}_R$. On being absorbed at the boundary $\partial \hat{\Omega}_R$ ($\mathcal{P}_N = \sigma$) the system of $N$ molecules reacts. Since the system is very large compared to $\sigma$, absorption of states on $\partial \hat{\Omega}_R$ causes a small perturbation within the bulk distribution. Thus, it is expected that a pseudo-steady state for  Equation (\ref{Nparticle_reduced}) is reached quickly. The steady state is given by
\begin{equation}\label{goveqn}
0 = \left[\sum_{i=2}^{N}
\hat{D}_i  \hat{\nabla}_i^2  \right] p,
\end{equation}
where $p( \boldsymbol{\eta}\in \partial\hat{\Omega}_R) = 0$ and $p$ sufficiently far from the origin should be equal to the unperturbed probability density $\lim_{\{\boldsymbol{\eta}\}_2^N\rightarrow \infty} p(\{\boldsymbol{\eta}\}_2^N,t) = p_\infty = 1/V^{N-1}$.

In order to significantly simplify the analysis, we shall use the following non-dimensional quantities for the metric of the state vectors $\|\boldsymbol{\eta}_i\|$, time $t$ and subsequent reaction rate $k$.
\begin{equation}\label{transforms}
r_i = \sqrt{\frac{\Delta_N}{\hat{D}_i}}\frac{||\boldsymbol{\eta}_i ||}{\sigma}, \ \ \tau = \frac{\Delta_N t}{\sigma^{2}},  \ \ \kappa = \frac{k\sigma^{2}}{\Delta_N V^{N-1}}.
\end{equation}

Here, the rescaling of space is designed so that the region for reaction $\hat{\Omega}_R$ becomes a unit hypersphere. Comparing the spatial rescaling (\ref{transforms}) with (\ref{goveqn}), the diffusion of the state vector $\boldsymbol{\eta}$ becomes isotropic in the non-dimensional coordinates. The rescaling of time is designed to make the isotropic diffusion constant equal to one in the non-dimensional coordinates.  
The non-dimensionalisation requires the renormalisation of the probability density which is rescaled with respect to the spatial dilations. 

\begin{equation}\label{pchange}
p' = \sigma^{3(N-1)}\mathcal{D}p,
\end{equation} 
where
\begin{equation}
\mathcal{D} = \left[\prod_{i=2}^{N}\left(\frac{\hat{D}_i}{\Delta_N}\right)^{3/2}\right].
\end{equation}
As a result, the non-dimensionalised probability density far away from $\hat{\Omega}_R$ is given by
\begin{equation}\label{pinf}
p'_\infty = \mathcal{D}\left(\frac{\sigma^{3}}{V}\right)^{N-1}.
\end{equation} 
Note that for the remainder of this document, in the interest of notational simplicity, we shall drop the $'$ notation for $p'$.

Writing the non-dimensional form of Equation (\ref{goveqn}), we find that the partial differential equation in $3(N-1)$ coordinates has radial symmetry. That is, by defining the radius in the normalised (non-dimensional) state space as  $r = \sqrt{\sum_{i=1}^{N-1}r_i^2}$ (note that this is actually the non-dimensionalisation of the proximity $\mathcal{P}_N$ of the molecules)  we find that the non-dimensional form of Equation (\ref{goveqn}) becomes an ordinary differential equation in $r$. Note that $r=1$ corresponds to $\partial \hat{\Omega}_R$ and $r\sigma = \mathcal{P}_N$ is the proximity of the reactants. Equation (\ref{goveqn}) simplifies to the ordinary differential equation
\begin{equation}\label{1deq}
\frac{d^2p}{dr^2} + \frac{(2\alpha + 1)}{r}\frac{dp}{dr} = 0,
\end{equation}
where $\alpha = (3N-5)/2$ is a parameter that appears frequently so has been given its own symbol. Equation (\ref{1deq}) is subject to the conditions $p(1) = 0$ and $\lim_{r\rightarrow \infty} p(r) = p'_\infty$ (see Equation (\ref{pinf})).
The general solution to (\ref{1deq}) can be written in the following form
\begin{equation}\label{1dsol}
 p(r) = a_1 + \frac{a_2}{r^{2\alpha}},
\end{equation}
where $a_1$ and $a_2$ are arbitrary constants. Using the boundary conditions at $r=1$ and $r\rightarrow \infty$, we obtain
\begin{equation}\label{1dsols}
  p(r) = p'_\infty \left( 1 - \frac{1}{r^{2\alpha}} \right) = \mathcal{D}\left(\frac{\sigma^{3}}{V}\right)^{2/3(\alpha+1)}\left( 1 - \frac{1}{r^{2\alpha}} \right)
\end{equation}

Matching the non-dimensional reaction rate $\kappa$ with the correct reaction radius $\sigma$ requires that $\kappa$ be equal to the total flux of the probability density over the hypersurface of the unit sphere defined by $r=1$. That is,
\begin{equation}\label{condition}
\kappa = \left. S_{2(\alpha+1)}\frac{dp}{dr} \right|_{r=1},
\end{equation}
where $S_{m} = m\pi^{m/2}/\Gamma(m/2+1)$ is the surface area of a unit $m$-dimensional  sphere. Re-dimensionalising Equation (\ref{condition}) and using (\ref{1dsols}) gives
\begin{equation}\label{reactionrate2}
k = \frac{4\pi^{\alpha+1}\sigma^{2\alpha}\Delta_N\mathcal{D}}{\Gamma(\alpha)},
\end{equation}
which simplifies to Equation (\ref{gensmol}).
Rearranging for $\sigma$ gives
\begin{equation}\label{multismol}
\sigma = \left(  \frac{k \Gamma(\alpha)}{4\pi^{\alpha+1}\Delta_N\mathcal{D}}\right)^{1/(2\alpha)}.
\end{equation}
The well known bimolecular Smoluchowski reaction radius can be found in the special case of $N=2$ ($\alpha=1/2$)
\begin{equation}
\sigma = \frac{k}{4\pi\left(D_1 + D_2 \right)}.
\end{equation} 

Since the theoretical framework presented here is a higher order analogue of Smoluchowski's original theory, many of the analytic results that have proceeded Smoluchowski may easily be extended to this theoretical framework.

In 1976, Masao Doi presented a modified model for chemical reactions \cite{Doi1,Doi2}. According to Doi's model, reactions occur with a constant rate $\lambda$ inside of the reaction radius rather than instantaneously like the Smoluchowski model. For bimolecular reactions, the Doi reaction radius $\sigma_\lambda$ differs from the Smoluchowski reaction radius $\sigma$ and is related to the reaction rate $k$ and the rate $\lambda$ through the following relationship
\begin{equation}
	k = 4\pi\hat{D}\left( \sigma_\lambda - \sqrt{\frac{\hat{D}}{\lambda}}\mathrm{tanh}\left[\sigma_\lambda  \sqrt{\frac{\lambda}{\hat{D}}}\right]\right).
\end{equation}
The derivation of this equation can be found in Erban and Chapman's paper \cite{erbanchapman}.  Generalising the Doi model to $N$-th order reactions, we may register a reaction at a rate $\lambda$ whenever the reactant proximities $\mathcal{P}_N \leq \sigma_\lambda$. Using the same analysis as that used by Erban and Chapman, the generalised Doi reaction radius $\sigma_\lambda$ for $N$-th order reactions as defined implicitly by the reaction rate $k$ and rate $\lambda$ may be easily derived.
\begin{equation}\label{reactionrate}
k = \frac{4\pi^{\alpha+1}\sigma_\lambda^{2\alpha}\Delta_N\mathcal{D}}{\Gamma(\alpha)}    \left[1-\frac{2\alpha \mathrm{I}_{\alpha}\left(\sigma_\lambda\sqrt{\frac{\lambda}{\Delta_N}}\right)}{\alpha\mathrm{I}_{\alpha}\left(\sigma_\lambda\sqrt{\frac{\lambda}{\Delta_N}}\right) + \left(\sigma_\lambda\sqrt{\frac{\lambda}{\Delta_N}}\right)\mathrm{I'}_{\alpha}\left(\sigma_\lambda\sqrt{\frac{\lambda}{\Delta_N}}\right)} \right],                           
\end{equation}
where $I_\alpha$ is the modified Bessel function of the first kind and order $\alpha$ and the $'$ indicates a derivative with respect to the argument of the function.

\section{Numerical simulation of high order reactions}

Modern approaches to the simulation of diffusion-limited reactions using a Smoluchowski framework are subdivided into two main approaches. The first approach, known collectively as the `event-driven' approach, involves solving first passage times for `contact' events (when molecules come within the Smoluchowski radius). Examples of algorithms which use this approach include the enhanced Green's function reaction dynamics (eGFRD \cite{eGFRD}) and the first passage kinetic Monte-Carlo (FPKMC \cite{FPKMC}) algorithms. These algorithms calculate precise moments that the proximity of reactants touch the Smoluchowski radius. Reaction events occur at these simulated moments in time and the algorithm proceeds asynchronously through time to coincide with these events. The second approach, known as `time-driven', progresses through time using prescribed finite timesteps. At each timestep, a decision is made about whether or not `contact' events have occurred. Timesteps can be chosen adaptively (such as in the original Green's function reaction dynamics, GFRD, algorithm \cite{GFRD}) or be held constant.

A popular constant timestep algorithm and software developed using a Smoluchowski framework was published in 2004 by Andrews and Bray \cite{Andrews2004}. Their algorithm, due to its simplicity, accuracy and usefulness for a wide array of applications, has been cited over 300 times. Their algorithm is implemented in the popular software Smoldyn (named for its simulation of \textit{Smol}uchowski \textit{dyn}amics). The Smoldyn package simulates diffusion and bimolecular reaction with single molecule detail. Molecule coordinates $\mathbf{x}_i(t)$ for each molecule $i$ are stored and updated at discrete times separated by a prescribed timestep $\Delta t$. Updates for the positions at each timestep are generated randomly according to the relationship
\begin{equation}\label{update}
\mathbf{x}_i(t+\Delta t) = \mathbf{x}_i(t) + \sqrt{2D_{i}\Delta t}\boldsymbol{\xi},
\end{equation}  
where $D_{i}$ is the molecule-specific diffusion constant and $\boldsymbol{\xi} = (\xi_1,\xi_2,\xi_3)$ is a vector of independent, normally-distributed, random numbers with unit variance and zero mean re-sampled for each timestep. At each timestep, if a pair of reactant molecules are within the corresponding Smoluchowski radius, a reaction is performed. The Smoluchowski radius $\sigma_{\Delta t}$ that needs to be used in the algorithm deviates from $\sigma$ defined in Equation (\ref{smol}). The deviation from the theoretical value occurs due to the introduction of discrete timesteps rather than continuous time. That is, whilst Equation (\ref{update}) exactly simulates diffusion, it does not account for instantaneous reaction at the Smoluchowski radius between reactants. Instead, molecules may artificially skip reaction by travelling in and out of the Smoluchowski condition within one timestep. As such, the radius $\sigma_{\Delta t}$ that is used in practise to correct for these reaction losses is slightly larger than the continuous-time theoretical Smoluchowski radius $\sigma$. The calculation of $\sigma_{\Delta t}$ in Smoldyn is described in their original manuscript \cite{Andrews2004}. For high order reactions, computation of $\sigma_{\Delta t}$ may be described using the presented framework and the same approach to that used in \cite{Andrews2004}. How this approach is generalised to the presented framework is discussed in detail later in this Section. As the algorithms in this section are generalisations of the algorithms first presented in Andrews and Bray's 2004 paper \cite{Andrews2004}, interested readers are directed to this reference for a discussion of the numerical considerations of this algorithm and its convergence properties.

Simulation of reaction-diffusion processes with high order reactions at the detail of individual molecules can be achieved with the algorithm outlined in Table \ref{alg:complete}. This algorithm is based on the Smoldyn time-driven algorithm using the generalised Smoluchowski framework presented in this manuscript and is the algorithm that is used for running the numerical tests in Section \ref{numericalsect}.

\begin{table}[h]
    \begin{enumerate}
    \setcounter{enumi}{0}
       \item[\mbox{[S.1]}] Define the domain and molecule initial positions $\mathbf{x}_i(0)$ for each molecule $i$. Decide on the temporal resolution of the simulation in timesteps $\Delta t$. Set the initial time $t=0$.
       \item[\mbox{[S.2]}] Calculate and store the scaled reaction radii $\sigma_{\Delta t}/\sqrt{\Delta_N}$ using the numerical technique described in the Section \ref{numalg} for all multimolecular reactions. 
       \item[\mbox{[S.3]}] Update the positions $\mathbf{x}_i(t+\Delta t)$ of each molecule $i$ using Equation (\ref{update}). If necessary, implement boundary conditions (see Ref. \cite{Andrews2004} for details).
       \item[\mbox{[S.4]}] For each possible reaction, find all combinations of reactants that have a scaled proximity $\mathcal{P}_N/\sqrt{\Delta_N}$ (defined by Equation (\ref{proximity})) less than the scaled reaction radius $\sigma_{\Delta t}/\sqrt{\Delta_N}$. Perform these reactions. Update the current time $t:=t+\Delta t$.
       \item[\mbox{[S.5]}] Repeat Steps [S.3] and [S.4] until the desired end of the simulation.
    \end{enumerate}
    \caption{Algorithm for the simulation of reaction-diffusion processes with high order reactions at the detail of individual molecules.}
    \label{alg:complete}
    \end{table}

\subsection{Numerical reaction rate $k$ using a reaction radius of $\sigma_{\Delta t}$} \label{numalg}

For numerical simulations that test for the reaction condition at discrete times separated by timesteps of duration $\Delta t$, we shall show how the reaction radius needs to be modified. The reaction-diffusion algorithm outlined in Table \ref{alg:complete} has two main steps in order to evolve through time and correctly sample reaction events. These are:

\begin{itemize}
\item[i] Update molecule positions using the formulae
$$\mathbf{x}_{i;j}(t+\Delta t) = \mathbf{x}_{i;j}(t) + \sqrt{2D_i\Delta t}\xi_j,$$ where $\mathbf{x}_{i;j}$ is the $j$-th component of the position vector $\mathbf{x}_{i}$ and $\xi_j$ is a unit variance, normally distributed random number.
\item[ii] If the reaction condition given in Equation (\ref{R}) is satisfied between $N$ reactants (using the numerical reaction radius $\sigma_{\Delta t}$), then a reaction is implemented.
\end{itemize}

In step [i] of the simulation algorithm the non-dimensional proximity $r$ of the system of $N$ particles evolves according to free diffusion. Consider the scaled probability density for the proximity $r$ given by $g(r) = p(r)/p'_\infty$. After non-dimensionalising the timestep $\Delta t = \sigma_{\Delta t}^2 \Delta \tau / \hat{D}_2$, the distribution $g(r,\tau+\Delta \tau)$ at the end of step [i] can be found from the initial distribution $g(r,\tau)$ at the start of step [i] using the governing equation
\begin{equation}\label{goveqg}
\frac{\partial g}{\partial \tau} = \frac{1}{r^{2\alpha+1}}\frac{\partial}{\partial r}\left( r^{2\alpha+1}\frac{\partial g}{\partial r}\right).
\end{equation}
 For even values of $N$ it is possible to find analytic Green's functions to Equation (\ref{goveqg}) which allows for direct numerical calculation of $g(r,\tau+\Delta \tau)$ from $g(r,\tau)$. However, this is difficult for odd values of $N$. In order to calculate the reaction rate that coincides with the reaction radius $\sigma_{\Delta t}$, we therefore solve Equation (\ref{goveqg}) numerically using a simple forward Euler scheme from $\tau$ to $\tau+\Delta \tau$. We implement a zero derivative boundary condition at the origin. The spatial domain is numerically truncated for some large $r=R \gg 1$. To determine the adaptive boundary condition at $r=R$, $g(r,\tau)$ is fitted to the form $(1+a(\tau)/r^{2\alpha})$ which is known to be accurate far away from $r=1$ according to Equation (\ref{1dsol}): $a$ is found by least squares fitting at each timestep update of the PDE using the final 10\% of the numerically stored $g(r,\tau)$. The boundary condition that is implemented at $r=R$ is $g_r(R,\tau) = -2\alpha a(\tau)/r^{2\alpha+1}$ given the current fitted value for $a(\tau)$.
 
The reaction probability per timestep consistent with the macroscopic rate of reaction $\kappa\Delta \tau$ is given by the total probability that the molecule will react at step [ii] when $g(r,\tau)$ has reached pseudo-steady state between consecutive timesteps.
\begin{equation}\label{reducedRR}
\mathcal{K}=\frac{ 2(\alpha + 1)\kappa\Delta \tau}{S_{2(\alpha + 1)}p'_\infty} = \frac{2(\alpha + 1)k\Delta t}{\mathcal{D}S_{2(\alpha + 1)}\sigma_{\Delta t}^{3(N-1)}} = \lim_{\tau\rightarrow\infty} \int_0^1 2(\alpha + 1) r^{2\alpha + 1} g \ \mathrm{d}r.
\end{equation}

We shall refer to $\mathcal{K}$ as the \textit{reduced reaction rate} (to match the terminology given in \cite{Andrews2004}). It should be noted that we deviate slightly from the definition of $\mathcal{K}$ given in \cite{Andrews2004} by a factor of  $S_{2(\alpha + 1)}/[2(\alpha + 1)]$ (the volume of the unit $[3(N-1)]$-dimensional sphere) so that, in the large $\Delta t$ limit, the reduced reaction rate is 1 for any value of $N$. The reduced reaction rate can be thought of as the ratio of the probability for a reaction per time step in pseudo-equilibrium to the probability for a reaction per time step if $p=p_\infty$ everywhere within $\partial\hat{\Omega}_R$ (i.e $0<r\leq 1$). Importantly, to reach the steady state in the limit as $\tau\rightarrow \infty$ in Equation (\ref{reducedRR}), $g(r)$ should be set to zero for $0<r<1$ after each iteration of the timestep $\Delta \tau$. Setting $g(r)=0$ for $0<r<1$, at these discrete timesteps mimics the effect of these states undergoing reaction in step [ii] of the simulation algorithm.   

To find $k$ given a reaction radius of $\sigma_{\Delta t}$ and timestep $\Delta t$, we first check if $\Delta t \ll \frac{\sigma_{\Delta t}^2}{2\Delta_N}$. In this case, $\sigma_{\Delta t} \sim \sigma$ and we may use the theory outlined in Section \ref{derivtheory}. Alternatively, if  $\Delta t \gg \frac{\sigma_{\Delta t}^2}{2\Delta_N}$, then $\mathcal{K} \approx 1$ and $k$ can be found from Equation (\ref{reducedRR}). The following numerical steps are taken for $\Delta t \sim \frac{\sigma(k)^2}{2\Delta_N}$.

\begin{itemize}
\item[i] Find the non-dimensional timestep using $\sigma_{\Delta t}$ and $\Delta t$: $\Delta \tau = \Delta t \hat{D}_2/\sigma^2_{\Delta t}$.
\item[ii] Initialize $g(r,0)=1$ for $0<r<R$. Set $\tau = 0$. $R$ is chosen to be sufficiently far from the reacting radius $r=1$ (for the simulations in this manuscript, we chose $R=10$). Lattice points for storing values of $g$ should be separated by $\delta r\ll \sqrt{\delta \tau}$, where $\delta \tau$ is the time discretisation of the PDE (\ref{goveqg}). Reducing $\delta r$ monotonically increases accuracy of the numerical integration involved in the calculation of the reduced reaction rate (Equation (\ref{reducedRR})). For the data generated in this manuscript we found it sufficient to set $\delta r = 0.02$ and $\delta \tau = 4\times 10^{-6}$.
\item[iii] For $r\sim R$, fit the current distribution of $g(r)$ to a functional form $1+a/r^{2\alpha}$ to find $a$. The error that is generated as a result of this imposed functional form is controlled by the truncation parameter $R$. For large enough $R$, this functional form is asymptotically exact \cite{Andrews2004}. In this manuscript, $g(r)$ over the final 10\% of the numerically stored domain was used to find $a$ using least squares fitting.
\item[iv] Numerically evolve Equation (\ref{goveqg}) from $\tau$ to $\tau +\Delta \tau$ remembering to use recalculated fitted values for $a$ in the boundary condition at $r=R$.
\item[v] Calculate the putative reduced reaction rate, $\mathcal{K}$, for the current timestep, $\Delta \tau$, using the integral in Equation (\ref{reducedRR}). This integral should be solved numerically. The trapezoidal rule was used for this manuscript.
\item[vi] If the putative reduced reaction rate has converged (differs with an acceptable relative variation between timesteps), the reaction rate, $k$, can be found from $\mathcal{K}$ in Equation (\ref{reducedRR}). In this manuscript we accepted convergence if $\mathcal{K}$ differed by less than 1 part in $10^5$ over two consecutive timesteps. Otherwise, set $g(r)=0$ for $0<r<1$, update the current time $\tau:=\tau + \Delta \tau$ and repeat from step [iv] until convergence.  
\end{itemize}

\subsection{Numerical reaction radius $\sigma_{\Delta t}$ for a prescribed reaction rate $k$} \label{numalg2}
 It is common to be presented with the inverse problem of finding the reaction radius $\sigma_{\Delta t}$ given the diffusion-limited reaction rate $k$. Whilst it possible write a `guess and check' algorithm for finding $\sigma_{\Delta t}$ from $k$ using the algorithm in the previous section, here we use a look up table since finding $k$ using the previous algorithm can be computationally time consuming. Let us define $s'$ as the RMS displacement of the state in each dimension of non-dimensionalised state space per timestep $\Delta \tau$  (notation consistent with Andrews and Bray \cite{Andrews2004}), 
\begin{equation}\label{sdash}
s'=\sqrt{2\Delta \tau}=\frac{\sqrt{2 \Delta_N \Delta t}}{\sigma_{\Delta t}}.
\end{equation}
 Using the algorithm in the previous section, we first find and tabulate $\mathcal{K}$ for $s' = \exp\left(\delta\right)$, where $\delta = -3,-2.8,-2.6,\ldots,3$. Whilst $\mathcal{K}$ can be found for $\exp(-3)<s'<\exp(3)$ by interpolation of the tabulated data, we use known forms of $\mathcal{K}$ in the case of large $s'$ and small $s'$.
 
  For large $s'$ ($\Delta t\rightarrow \infty$), $g(r,\tau) = 1$ inside the integrand on the RHS of Equation (\ref{reducedRR}) and therefore 
\begin{equation}\label{largedt}
\mathcal{K} = 1.
\end{equation}
For small $s'$ ($\Delta t\rightarrow 0$), the Smoluchowski result is accurate. Therefore, using Equations (\ref{reactionrate2}), (\ref{reducedRR}) and (\ref{sdash}) it is possible to show that
\begin{equation}\label{smalldt1}
\mathcal{K} = 2\alpha\left(\alpha + 1 \right) s'^2.
\end{equation}
The following root bracketing algorithm used by Andrews and Bray \cite{Andrews2004} can be implemented to find the reaction radius $\sigma_{\Delta t}$ as a result of a reaction rate $k$ for the finite $\Delta t$ timestep Smoluchowski model.

\begin{itemize}
\item[i] Start with an initial guess for the reaction radius $\sigma_{\mathrm{guess}} = \sqrt{2 \Delta_N \Delta t}$ ($s'_{\mathrm{guess}} = 1$ from Equation (\ref{sdash})). Set a minimum bound for $\sigma_{\Delta t}$: $\sigma_{\mathrm{min}} = 0$.
\item[ii] Find the reduced reaction rate $\mathcal{K}_{\mathrm{guess}}$ as a result of the current guess for $s'_{\mathrm{guess}}$. $\mathcal{K}_{\mathrm{guess}}$ is found from using a cubic Lagrange interpolating polynomial applied to the data in the table as a function of $\delta = \mathrm{ln}(s')$. For data outside of the range of the table ($s' > \exp(3)$ and $s' < \exp(-3)$) the analytical formulas (Equations (\ref{largedt}) and (\ref{smalldt1}) respectively) may be used.
\item[iii] Find $k_{\mathrm{guess}}$ from $\mathcal{K}_{\mathrm{guess}}$ and $\sigma_{\mathrm{guess}}$ using (\ref{reducedRR}).
\item[iv] If $k_{\mathrm{guess}} < k$, update the minimum bound on the reaction radius $\sigma_{\mathrm{min}} := \sigma_{\mathrm{guess}}$ and double the guess for the reaction radius $\sigma_{\mathrm{guess}} := 2\sigma_{\mathrm{guess}}$ (subsequently halving $s'_{\mathrm{guess}}$). 
\item[v] Repeat steps [ii], [iii] and [iv] until $k_{\mathrm{guess}} \geq k$ and define $\Delta \sigma = \sigma_{\mathrm{guess}}-\sigma_{\mathrm{min}}$, a known bracket interval length for the value of $\sigma_{\Delta t}$.
\item[vi]  Split the difference between the minimum, $\sigma_{\mathrm{min}}$, and maximum, $\sigma_{\mathrm{min}}+\Delta \sigma$ reaction radii for a new guess $\sigma_{\mathrm{guess}} := \sigma_{\mathrm{min}} + \Delta \sigma/2$ and find the subsequent $s'_{\mathrm{guess}}$ using Equation (\ref{sdash}). 
\item[vii] Repeat steps [ii], [iii].
\item[viii] Whether or not $k_{\mathrm{guess}} > k$ or $k_{\mathrm{guess}} < k$, the interval between maximum and minimum reaction radii is halved: $\Delta \sigma:=\Delta \sigma/2$. If $k_{\mathrm{guess}} > k$, update the minimum bound on the reaction radius $\sigma_{\mathrm{min}} := \sigma_{\mathrm{guess}}$. 
\item[ix] Repeat steps [vi]-[viii]. For each repeat, the uncertainty in the guess, $\Delta \sigma$, is halved and linear convergence of the guess towards the true reaction rate is achieved. The author finds it sufficient to repeat 15 times.
\item[x] Set $\sigma_{\Delta t} = \sigma_{\mathrm{min}} + \Delta \sigma/2$.
\end{itemize}

\begin{figure}[!ht]
\includegraphics[width=1\columnwidth]{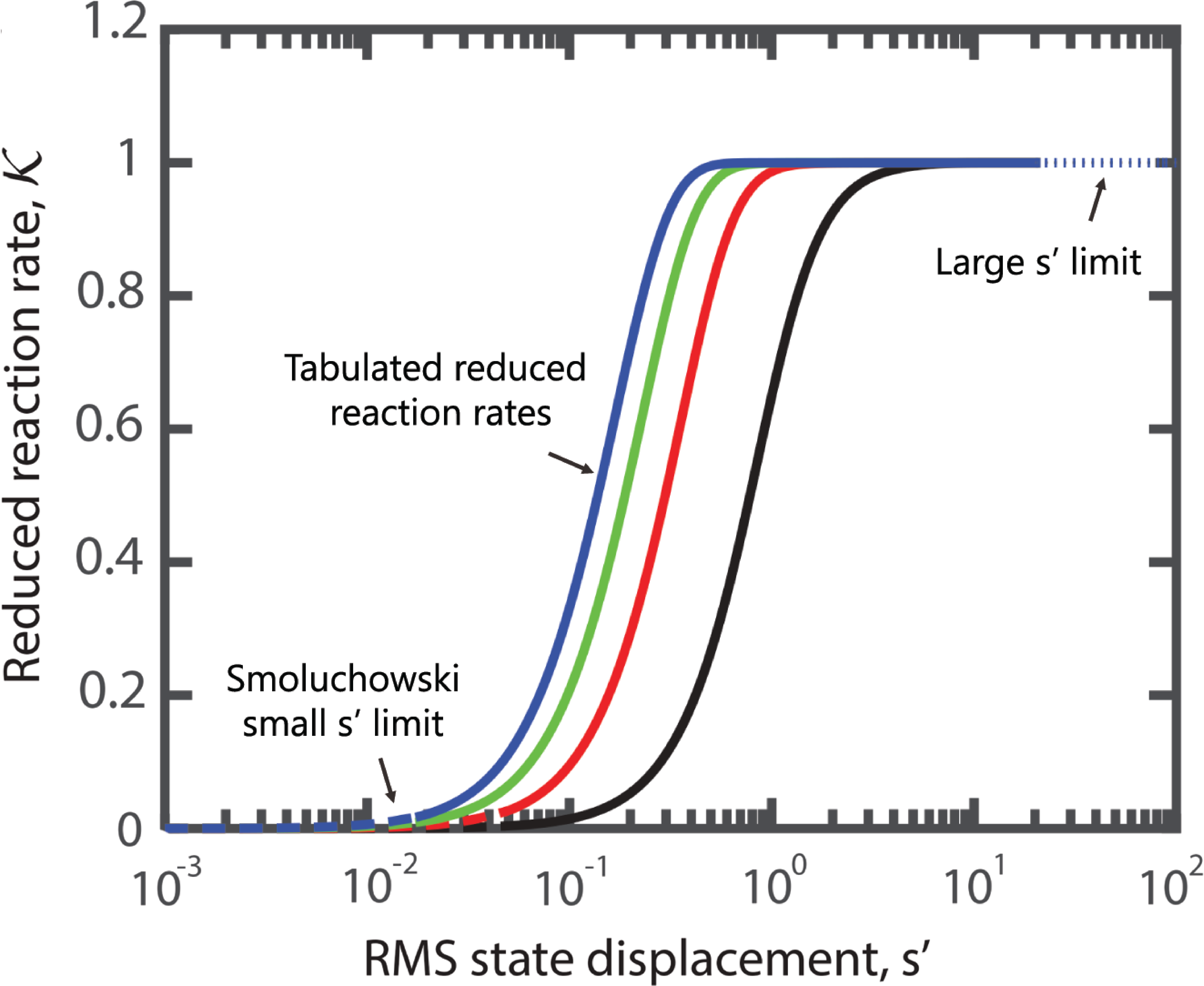}
\caption{A plot of reduced reaction rate $\mathcal{K}$ versus the RMS displacement $s'$ for reactions of order 2 (black), 3 (red), 4 (green) and 5 (blue). The solid portions of these curves were calculated using cubic interpolating polynomials }\label{redreactionrate_img}
\end{figure}

Figure \ref{redreactionrate_img} shows how the reduced reaction rate $\mathcal{K}$ depends on the RMS displacement $s'$. The interface between the transient tabulated reduced reaction rate and the asymptotic limits (\ref{largedt}) and (\ref{smalldt1}) are smooth indicating that the numerical algorithm which generated the tabulated data is converging to the correct reaction rate.

\section{Numerical tests and discussion}\label{numericalsect}
To validate the theory and explore the scientific implications of a Smoluchowski approach to multimolecular reactions we present three test problems. The first test problem demonstrates a simple well-mixed reaction of various orders. The test problem is designed to test if, indeed, the reaction rate $k$ is correctly determined by its associated value of the reaction radius $\sigma_{\Delta t}$. The second test problem explores the validity of the law of mass action as it is normally applied to multimolecular reactions. Finally, a simulation of a basic Turing system, which necessarily requires the implementation of multimolecular reactions, will be demonstrated using the stochastic algorithm in Table \ref{alg:complete}.  

\subsection{Well-mixed reactions}
\begin{figure}[!ht]
\includegraphics[width=1\columnwidth]{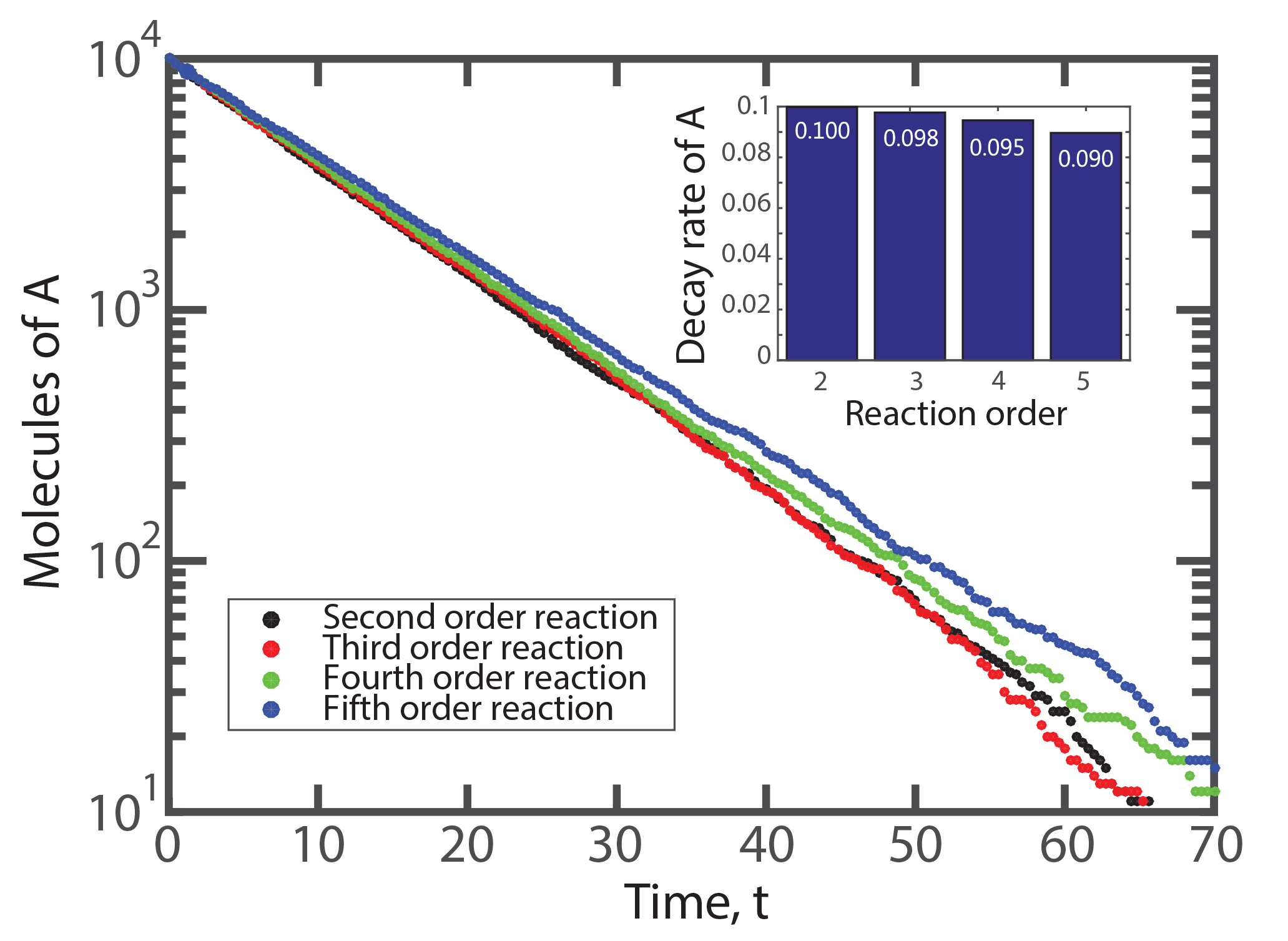}
\caption{Number of molecules of $A$ for each of the chemical systems (\ref{bireact})-(\ref{pentareact}) as a function of time $t$ using the generalised Smoluchowski framework. The initial number of molecules of $A$ is $10^4$, whilst other chemical species have $10^3$ molecules. For each of the simulations, the prescribed rate of decay of $A$ is the same ($k=0.1$) and compared with the simulation output. Whilst there is fairly good agreement between simulations, the simulated rate of decay is increasingly less than the theoretical rate $k$ as the order of the reaction increases. The simulated decay rate of $A$ is determined by non-linear least squares fit to an exponential decay model and presented in the subgraph in the top right corner. The timestep for each of the simulations was $\Delta t = 10^{-4}$ but plotted only after every $4000$ timesteps. The standard errors associated with the simulated reaction rates are very small (less than $5\times 10^{-6}$ in all four cases) and so are not shown with errorbars.}  \label{test3}
\end{figure}

In the first numerical test we simulate separately the following reactions of increasing order                                   
\begin{align}
A+B &\xrightarrow{k_2} B, \label{bireact} \\
A+B+C &\xrightarrow{k_3} B + C, \label{trireact}\\
A+B+C+D &\xrightarrow{k_4} B + C + D, \label{quadreact}\\
A+B+C+D+E &\xrightarrow{k_5} B + C + D + E.  \label{pentareact}
\end{align}
Since the generalised Smoluchowski theory simplifies to the classical theory for bimolecular reactions, the bimolecular reaction (\ref{bireact}) is simulated using the Smoldyn algorithm \cite{Andrews2004}. Each of the chemical reactions (\ref{bireact})-(\ref{pentareact}) result, in principle, in the simple exponential decay of $A$. This affords us the possibility to validate the numerical algorithm which simulates these reactions. Specifically, we wish to show that the average simulation rate of decay of $A$ matches that which we prescribe ($k_2$, $k_3$, $k_4$, and $k_5$ respectively). The molecular copy-numbers of $B$, $C$, $D$ and $E$ in each of the chemical reactions is conserved and for the sake of continuity between simulations of reactions with different orders were chosen to be equal ($10^3$ molecules each). The well-mixed domain is a dimensionless $1\times 1\times 1$ cube with periodic boundary conditions (to avoid boundary effects). In order to ensure the same prescribed rate of decay of $A$, which may be then compared for each type of simulated chemical reaction, the dimensionless reaction rates were chosen as $k = 10^{3}k_2 = 10^{6}k_3 = 10^{9}k_4 = 10^{12}k_5 = 0.1$ (the half-life of $A$ for each reaction is expected to be a dimensionless time of $10 \mathrm{ln}(2)$). The population of $A$ is initialised at $10^4$ molecules such that intrinsic simulation noise is as small as possible and only apparent at large times. Fig. \ref{test3} demonstrates that each simulation reproduces the correct prescribed reaction rate. As the order of the reaction rate is increased, the simulated reaction rate increasingly undervalues the theoretical reaction rate. This is because the density of reactant combinations significantly increases with reaction order and the Smoluchowski dilute limit becomes increasingly invalid. 

\subsection{Mass action versus Smoluchowski kinetics}
\begin{figure}[!ht]
\includegraphics[width=1\columnwidth]{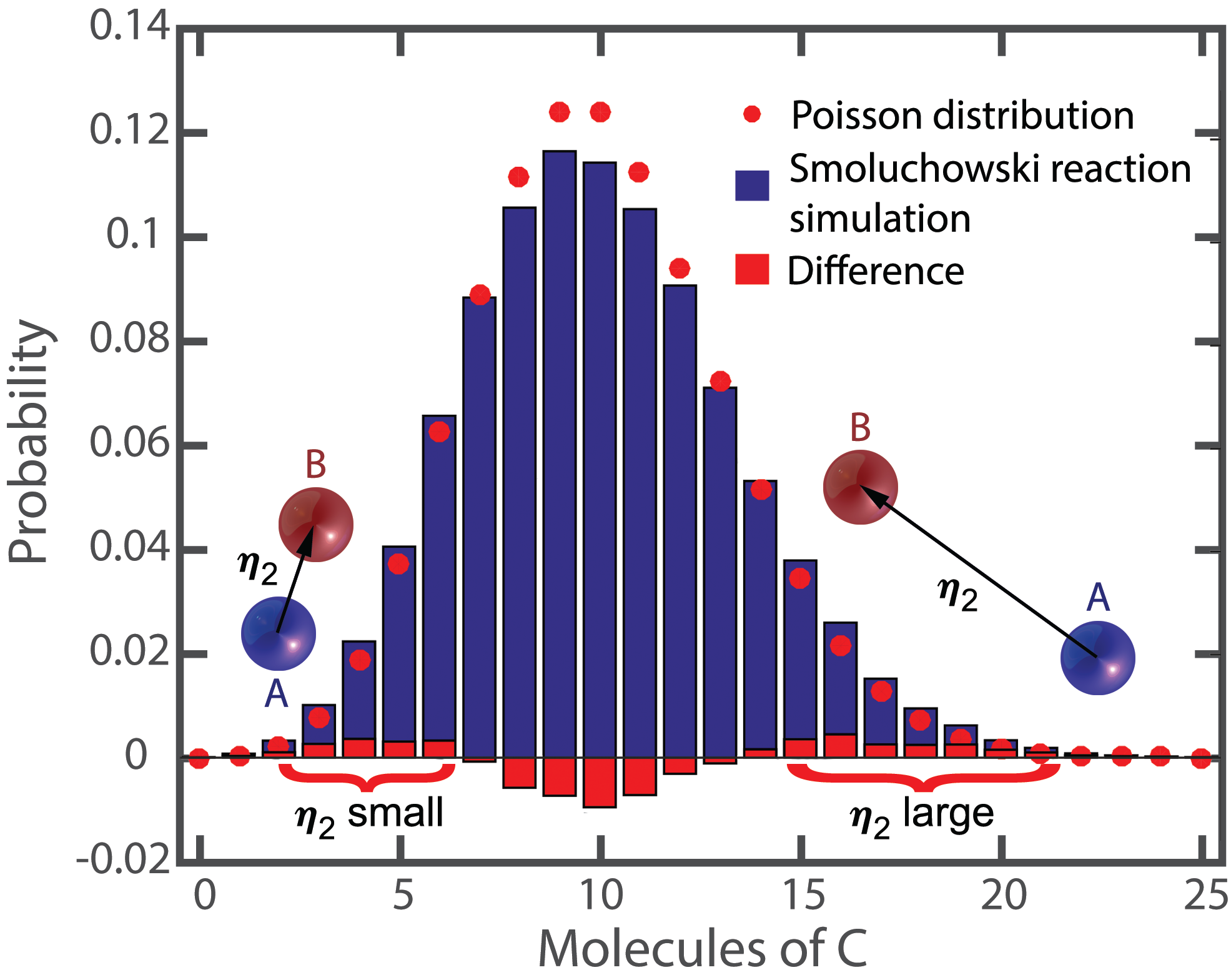}
\caption{Number of molecules of $C$ for the chemical system (\ref{massaction}) after a dimensionless time interval $t=18$. The blue bars represent the distribution sampled from $2\times 10^5$ simulations of the generalised Smoluchowski algorithm. The red dots represent the Poisson distribution predicted by the law of mass action. The red bars represent the difference between these two distributions indicating that Smoluchowski reaction kinetics tend to favour the two tails of the distribution. This is due to a the fact that molecules $A$ and $B$ are more likely to be near each other when the number of molecules of $C$ is small and at a distance when the number of molecules of $C$ is large. The maximum error associated with these histograms as a result of $2\times 10^5$ simulations is less than $10^{-3}$ and are therefore not shown on this figure. }  \label{test1}
\end{figure}

In this subsection we test the well-mixed chemical system
\begin{equation}\label{massaction}
A+B+C \xrightarrow{k_2} A+B, \quad \quad  \emptyset\xrightarrow{k_1} C,
\end{equation}
inside a unit cube with periodic boundary conditions (so that boundary effects do not contaminate the results of the simulations). Only one immortal molecule of both $A$ and $B$ are initially placed at random within the cube to minimise crowding and obtain the best possible reaction rate. We shall use non-dimensional rate parameters $k_1 = 1$ and $k_2 = 0.1$. This type of chemical system is often treated using the law of mass action. According to the law of mass action, at any moment in time, the propensity for the creation of a molecule of C is given by $k_1$ and the propensity for the trimolecular removal of a molecule of C is given by $k_2N_C$ where $N_C$ is the current number of molecules of C (the reader is reminded that $N_A=N_B=1$ for all time). Using the associated master equation for this birth-death process it is trivial to show that the steady state distribution of $N_C$ ought to be the Poisson distribution, $N_C \sim \mathrm{Pois}(k_1/k_2) = \mathrm{Pois}(10)$. To test if $A$, $B$ and $C$ obey mass action kinetics in a Smoluchowski model of trimolecular reactions we initialise $N_C$ molecules uniformly distributed over the unit volume. The number of molecules $N_C$ is sampled from the mass action steady state $N_C\sim\mathrm{Pois}(10)$. Choosing non-dimensional diffusion constants $D_A=D_B=D_C=1$, the chemical system is simulated using the generalised Smoluchowski kinetics and a timestep of $\Delta t = 5\times 10^{-4}$ for a non-dimensional duration of $t=20$. This simulation was repeated $2\times 10^5$ times and the distribution of $N_C$ was sampled. In Fig. \ref{test1} the distribution of $N_C$ after simulation (blue bars) is compared with the Poisson distribution with a mean of 10 (red dots). The difference between the Smoluchowski simulation and the Poisson (mass action) distributions is indicated using red bars. It is clear that the variance of the initial Poisson distribution has increased (from $10$ to $\sim 11.65$). The final number of molecules of C are more likely to be in the two tails of the distribution.     Because the Poisson distribution is skewed, the overall average number of molecules of C is also slightly increased from $10$ to $\sim 10.07$. 
The law of mass action assumes that the current rate of reaction depends on the current state of the chemical system and has no memory of past events. This assumption is broken for multimolecular diffusion-controlled reactions. Molecules of $C$ can only be removed if the molecules of $A$ and $B$ are sufficiently close. Given that a molecule of $C$ has just been removed, it is more likely that $A$ and $B$ are close ($\boldsymbol{\eta}_2$ small) and that other molecules of $C$ will be removed in the near future. Conversely, if it has been a significant time since the last removal of $C$ from the system, it is more likely that $A$ and $B$ are far apart ($\boldsymbol{\eta}_2$ large) and this reduces the probability of a molecule of $C$ being removed in the near future. Consequently, the number of molecules of $C$ linger at the two tails of the distribution for longer than they would under the assumptions of mass action. If the temporal correlation of $\boldsymbol{\eta}_2$ disperses more rapidly (by an increase in $D_A$ or $D_B$) the final distribution of the Smoluchowski reaction simulations more closely matches the Poisson distribution predicted by mass action. Conversely, the discrepancy is emphasised by decreasing $D_A$ and $D_B$.

\subsection{Simulation of Turing patterns}

\begin{figure}[!ht]
\includegraphics[width=1\columnwidth]{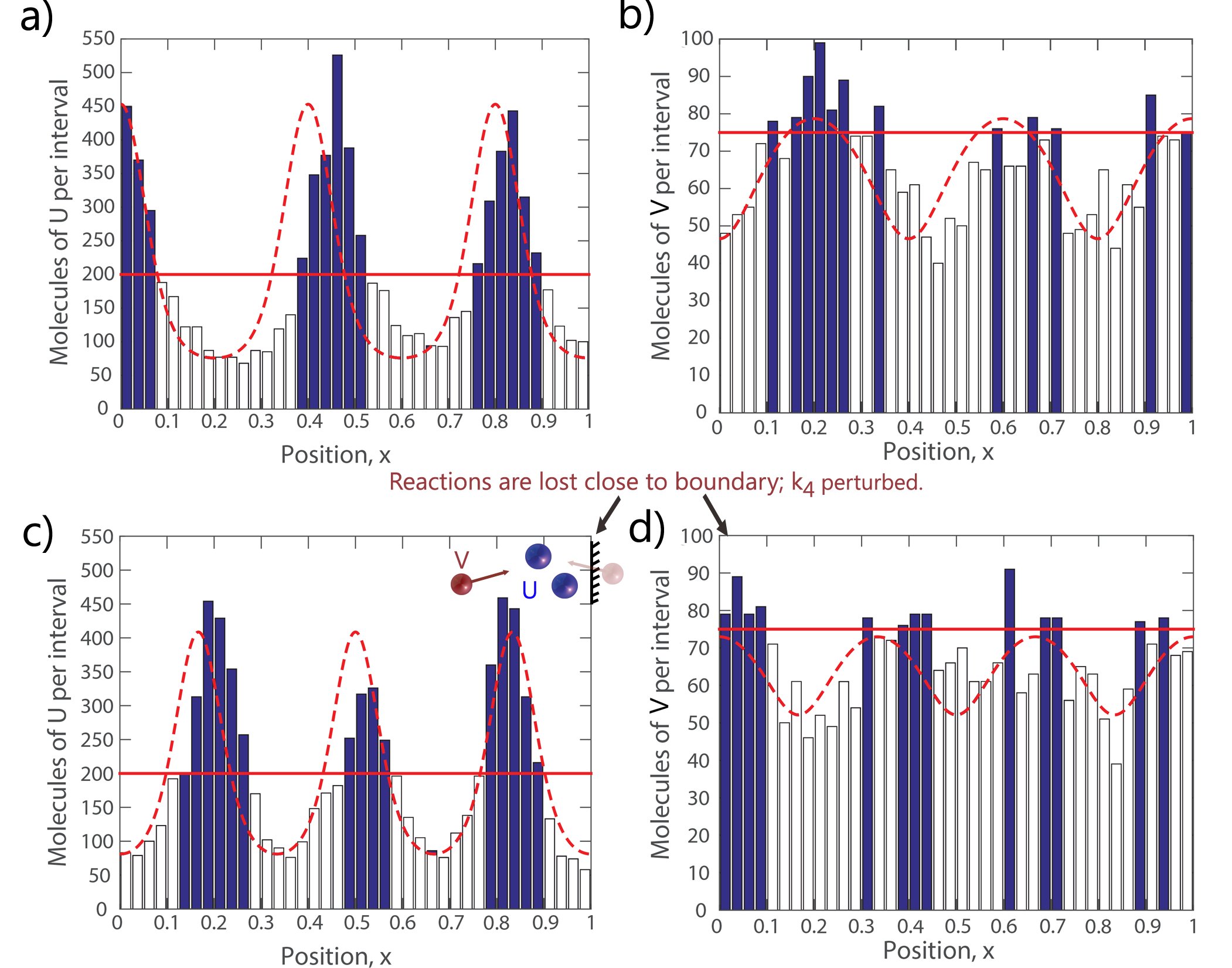}
\caption{Histogram plot of the number of molecules per interval of width $h=0.025$ at time $t=18$ for the chemical system in Equation (\ref{Turingsystem}). a) Molecules of U and b) molecules of V for the chemical system simulated using the classical spatial Gillespie algorithm as implemented by Cao and Erban \cite{Cao2014} on the regular lattice with spacing $h=0.025$. c) Molecules of U and d) molecules of V for the chemical system simulated using the generalised Smoluchowski framework with a timestep of $\Delta t = 3.125\times 10^{-5}$. The red solid line in each figure corresponds to the uniform unstable steady state of the PDE system (\ref{u}) and (\ref{v}) and the initial condition for the simulations. The red dashed lines are the steady state Turing solutions to the PDE system (\ref{u}) and (\ref{v}) solved using Matlab's PDEPE solver with a spatial resolution of $\delta x = 10^{-4}$. Bars coloured blue indicate molecular copy numbers greater than the uniform steady state so that the Turing pattern can be more easily identified. It should be noted that for the PDE solution in Figures \ref{Turingresults}c and \ref{Turingresults}d a perturbed heterogeneous form for $k_4(x)$ is used (see Equation (\ref{perturbation})) to mimic the boundary effects consistent with the diffusion-controlled reactions simulated by the Smoluchowski framework.}\label{Turingresults}
\end{figure}

In 1952, Alan Turing demonstrated a mathematical basis for how a system of reacting chemical species undergoing diffusion may spontaneously exhibit heterogeneity from an homogeneous initial distribution \cite{Turing}. The resultant heterogeneity in the chemical concentrations are known as Turing patterns and have since been observed widely in chemistry and biology \cite{Maini1997}. Often Turing patterns are modelled using partial differential equations (PDEs). At low concentrations, PDEs do not accurately represent the stochastic behaviour of a chemical system. In this subsection, we shall demonstrate that the generalised Smoluchowski framework is capable of simulating stochastic Turing patterns. We shall use the example chemical system 
\begin{equation}\label{Turingsystem}
\emptyset \underset{k_2}{\stackrel{k_1}{\rightleftharpoons}} U, \quad \quad   
\emptyset \xrightarrow{k_3} V, \quad \quad  2U+V\xrightarrow{k_4} 3U.
\end{equation}
This simple Turing system was first studied by J\"{u}rgen Schnakenberg \cite{schnakenberg} and subsequently it bears his name. The one-dimensional mean-field PDE description of this chemical system in the presence of diffusion is given by
\begin{align}\label{u}
\frac{\partial u}{\partial t} &= D_u \frac{\partial^2 u}{\partial x^2} + \bar{k}_1 -\bar{k}_2 u + \bar{k}_4u^2v, \\ \label{v}
\frac{\partial v}{\partial t} &= D_v \frac{\partial^2 v}{\partial x^2} + \bar{k}_3 - \bar{k}_4u^2v ,
\end{align}
where $u$ and $v$ are concentrations of $U$ and $V$ per unit length, the bar notation here denotes that the chemical rates are with respect to concentration per unit length and not volume and $D_u$ and $D_v$ are the Einstein diffusion constants for molecules of type $U$ and $V$ respectively.
A widely accepted approach to studying the stochastic nature of this chemical system in the presence of diffusion is with the use of well-mixed compartments and the spatial Gillespie algorithm (which solves exactly for realisations of the reaction-diffusion master equation). In their recent publication, Cao and Erban \cite{Cao2014} demonstrated that using such an approach to study Turing patterns can be difficult since the pattern may be affected by the choice of lattice which bring the results of the simulation into question. Cao and Erban specifically used the Schnakenberg system in (\ref{Turingsystem}) on a dimensionless 1D domain from 0 to 1. For this test problem, we shall use dimensionless parameters that were used in their paper: $\bar{k}_1 = 4\times 10^{3}$, $\bar{k}_2 = 2$, $\bar{k}_3 = 1.2\times 10^{4}$, $\bar{k}_4 = 6.25\times 10^{-8}$, $D_u = 10^{-3}$ and $D_v = 10^{-1}$. Since the Smoluchowski dynamics is written for three-dimensional spatial domains, we simulate the reaction-diffusion system (\ref{Turingsystem}) on a rectangular prism domain of volume $1\times h \times h$ where $h = 0.025$, inspired by the compartment-size used in Cao and Erban's paper, are the side lengths of the domain in the $y$ and $z$ directions. We implement periodic boundary conditions in the $y$ and $z$ direction and no flux boundary conditions at $x=0$ and $x=1$. Importantly, since we have translated the test problem from one dimension to three dimensions, the chemical reaction rates should be translated into volumetric reaction rates. That is, we use the translated reaction rates $k_1 = \bar{k}_1/h^2$, $k_2 = \bar{k}_2$, $k_3 = \bar{k}_3/h^2$ and $k_4 = \bar{k}_4h^4$. A timestep of $\Delta t = 3.125\times 10^{-5}$ and a uniform initial condition consistent with the unstable uniform steady state of the chemical system ($8000$ total molecules of $U$ and $3000$ total molecules of $V$) was used. Simulations were performed using both the classical spatial Gillespie algorithm from Cao and Erban \cite{Cao2014} and a generalised Smoluchowski simulation. The results of these simulations at $t=18$ can be seen in Fig. \ref{Turingresults}. The spatial Gillespie algorithm produced two and a half peaks of both $U$ and $V$ which is consistent with the PDE system (\ref{u}) and (\ref{v}). On the other hand, Smoluchowski simulations produced three distinct peaks (with a significantly reduced amplitude) in $U$ and peaks in $V$ between those of $U$. Simulations of multimolecular diffusion-controlled reactions in a thin layer of space near boundaries have been shown to exhibit a perturbed (reduced) reaction rate \cite{Andrews2015} (although this has never been reported experimentally). The reason for this is due to the reduction of total flux over the reaction radius due to the reduction of molecular flux over the boundary. Whilst a rigorous analysis of this effect is yet to be published, it can be shown that a very small perturbation in the value of $k_4$ near the boundary is capable of completely changing the steady state Turing pattern and that change is predicted by Smoluchowski simulations.  
In Figures \ref{Turingresults}c and \ref{Turingresults}d we compare the molecular distribution  of $U$ and $V$ as a result of Smoluchowski simulation with the steady state of the PDE system (\ref{u}) and (\ref{v}) where a perturbation to $k_4$ is introduced
\begin{equation}\label{perturbation}
k_4(x) = k_4 \left(1- \epsilon_1\mathrm{exp}\left( \frac{-x}{\epsilon_2}\right) - \epsilon_1\mathrm{exp}\left( \frac{-(1-x)}{\epsilon_2}\right) \right).
\end{equation}   
where $\epsilon_1 = 0.01$ and $\epsilon_2 = 10^{-3}$ are chosen to be conservatively small. This perturbation to $k_4$ was able to force a completely different Turing pattern. In order to test that the three-peaked Turing pattern was an attractor state for the Smoluchowski simulation (and not just an artefact of the initial conditions near the unstable homogeneous steady state), a Smoluchowski simulation was run until $t=18$ using the classical Gillespie algorithm distribution in Figures \ref{Turingresults}a and \ref{Turingresults}b as an initial condition. The Gillespie Turing pattern reliably evolved into the three-peaked Smoluchowski Turing pattern. On a final note, it can be seen fairly clearly in Figure \ref{Turingresults} that stochastic simulations, both the spatial Gillespie and the Smoluchowski simulations, do not generate peaks which coincide with the peaks of the PDE precisely. This discrepancy is due to intrinsic stochastic noise and not due to the errors in the approximations or implementations of the simulation algorithms.

\section{Conclusion}

In this paper we have extended the Smoluchowski theory to encompass diffusion-controlled reactions of any order in three dimensions. The reaction radius in this generalised framework represents the proximity of a set of reactants, bellow which, a reaction may occur. The definition of proximity in this context is quite deliberate. It is defined by Equation  (\ref{proximity}) and utilises the displacement of each successive reactant from the centre of diffusion of the previous reactants. Whilst the proximity is equivalent to the square root of the weighted mean square distance between all pairs of reactants, it is more convenient to use (\ref{proximity}) as a definition.  It can be shown that the Smoluchowski reaction radius should then be given implicitly by Equation (\ref{gensmol}). Using this mathematical framework, a number of interesting phenomena involving high order reactions were observed. High order reactions have a reduced reaction rate due to significantly increased density of reactant combinations. Furthermore, since  reactions of order three or higher require the proximity of two or more reactants before a final molecule can cause a reaction to occur, memory is implicit in these reaction-diffusion systems and subsequently this may cause mass action kinetics to be inaccurate. Finally, multimolecular diffusion controlled reactions have a reduced reaction rate near boundaries and this may cause heterogeneous reaction-diffusion patterns to behave differently than the classical reaction-diffusion PDE  would indicate.

\appendix
\section{Invariance of the generalised Smoluchowski theory under molecular relabelling}

One of the defining attributes of the generalised Smoluchowski theory, as presented in this manuscript, is that a molecular order must be defined before the proximity $\mathcal{P}_N$ may be calculated and compared with the reaction radius. This is because, according to Equation (\ref{proximity}), the proximity involves summing weighted contributions of square lengths, $||\boldsymbol{\eta}_i||^2$, and relative diffusion constants $\hat{D}_i$, which are both functions of the chosen molecular order.  Seemingly, $\mathcal{P}_N$ depends on the order in which you choose to label the molecules, however the choice of labelling was done arbitrarily and so should give the same result regardless of the order. Here we shall show that the definition for the proximity of $N$ molecules $\mathcal{P}_N$ as well as the reaction rate $k$ defined by a reaction radius $\sigma$ is invariant under the choice of molecule labelling.

The reaction rate $k$ is given by the reaction radius $\sigma$ according to Equation (\ref{gensmol})

\begin{equation}\label{gensmolappend}
k = \left[\prod_{i=2}^{N}\hat{D}_i^{3/2}\right] \frac{4\pi^{\alpha+1}}{\Gamma(\alpha)}\left(\frac{\sigma}{\sqrt{\Delta_N}} \right)^{2\alpha}.
\end{equation}

\begin{figure}[!ht]
\begin{center}
\includegraphics[width=0.5\columnwidth]{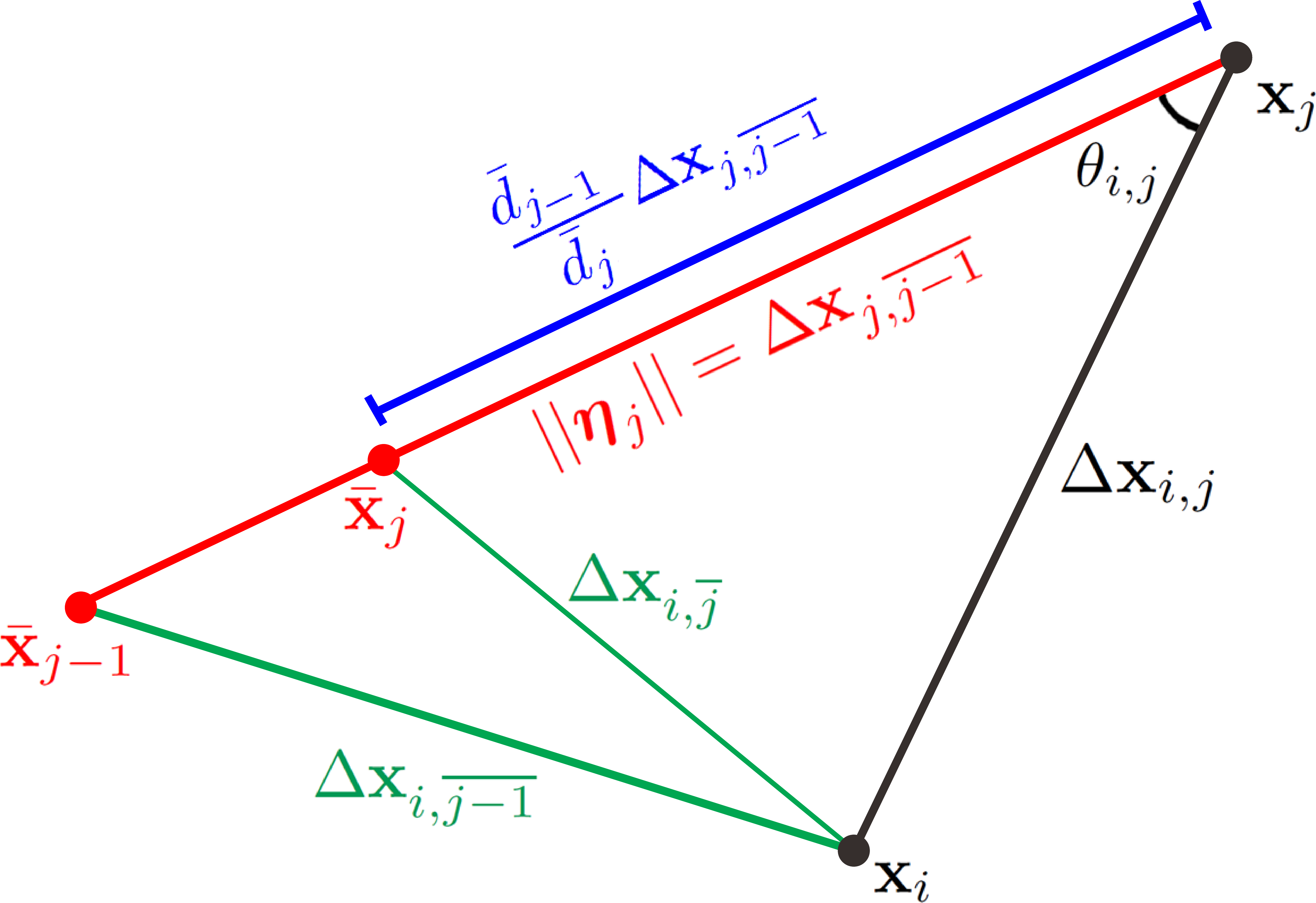}
\caption{Diagram of distances used to find Equation (\ref{triangleeqn}). The red line corresponds to a distance determined by a separation vector $||\boldsymbol{\eta}_j||$, the blue line is a portion of the red line and is determined by the relative diffusive weights at $\mathbf{x}_j$ and the centre of diffusion of the previous $j-1$ molecules, $\bar{\mathbf{x}}_{j-1}$. The two green lines do not, necessarily, correspond to any of the separation vectors but are distances that iteratively need to be simplified using Equation (\ref{triangleeqn}). The black line is a distance between two molecules.}\label{triangle}
\end{center}
\end{figure}

The diffusion constant for the $i$-th separation vector $\hat{D}_i$ depends on the labelling of the molecules, however, using Equations (\ref{Dbar}) and (\ref{realDbar}) we find that 
\begin{equation}
\hat{D}_i = \frac{1}{d_i} + \frac{1}{\bar{d}_{i-1}} = \frac{\overline{d}_{i}}{d_i\overline{d}_{i-1}}, \quad \text{ for } 2\geq i \geq N,
\end{equation}
where, for the sake of notational simplicity, we have used the notation $d_i = D_i^{-1}$ and $\overline{d}_i = \bar{D}_i^{-1} = \sum_{m=1}^i d_m$ ($d_i$ acts as the diffusive weight of the $i$th molecule and $\overline{d}_i$ the total diffusive weight of the first $i$ molecules used when calculating centres of diffusion). Thus (since $\overline{d}_1 = d_1$), 
\begin{equation}
\prod_{i=2}^N \hat{D}_i = \frac{\left(\prod_{i=2}^N \overline{d}_{i} \right)}{\left(\prod_{i=2}^N d_i \right)\left(\prod_{i=1}^{N-1}\overline{d}_{i-1}\right)} = \frac{\overline{d}_N}{\prod_{i=1}^N d_i} = \overline{d}_N \prod_{i=1}^N D_i
\end{equation}
 is independent of the labelling order. Furthermore, we have, from Equation (\ref{scalefact}),
 \begin{equation}\label{scalefact2}
\Delta_N = \frac{\sum_{i=1}^{N} D_i^{-1}}{\sum_{i>m} (D_i D_m)^{-1}}= \frac{\overline{d}_N}{\sum_{i>m} d_id_m},
\end{equation}
which is also independent of the labelling order. Therefore, according to the result (\ref{gensmolappend}), $k$ is independent of the labelling order if and only if $\sigma$ (which is nothing other than a particular proximity $\mathcal{P}_N$) is also independent of the labelling order. The measure of the proximity of $N$ reactants is given by Equation (\ref{proximity})
\begin{equation}\label{appendeqn}
\frac{\mathcal{P}_N^2}{\Delta_N} = \sum_{i=2}^N \frac{||\boldsymbol{\eta}_i||^2}{\hat{D}_i} = \sum_{i=2}^N \frac{d_i\overline{d}_{i-1}||\boldsymbol{\eta}_i||^2}{\overline{d}_i}.
\end{equation}
Since we already know $\Delta_N$ is independent of the molecular order, we wish to show that $\mathcal{P}_N^2/\Delta_N$ is independent of the molecular order. In order to show that this is the case, we need to write $||\boldsymbol{\eta}_i||$ in terms of relative molecular distances. We denote $\Delta \mathbf{x}_{i,j} = ||\mathbf{x}_i - \mathbf{x}_j||$ and $\Delta \mathbf{x}_{i,\bar{j}} = ||\mathbf{x}_i - \bar{\mathbf{x}}_j||$. Since $\Delta \mathbf{x}_{i,j} = \Delta \mathbf{x}_{j,i}$ and $\Delta \mathbf{x}_{i,\bar{j}} = \Delta \mathbf{x}_{\bar{j},i}$,  we shall only consider distances where the first index is larger than the second index. We wish to write all distances of the form $\Delta \mathbf{x}_{i,\bar{j}}$ in terms of relative molecular distances of the form $\Delta \mathbf{x}_{i,j}$ for some combinations of $i$ and $j$. Using Figure \ref{triangle} and applying the cosine rule twice, we can write an iterative identity for distances of the form $\Delta \mathbf{x}_{i,\bar{j}}$.

\begin{figure}[!ht]
\begin{center}
\includegraphics[width=0.5\columnwidth]{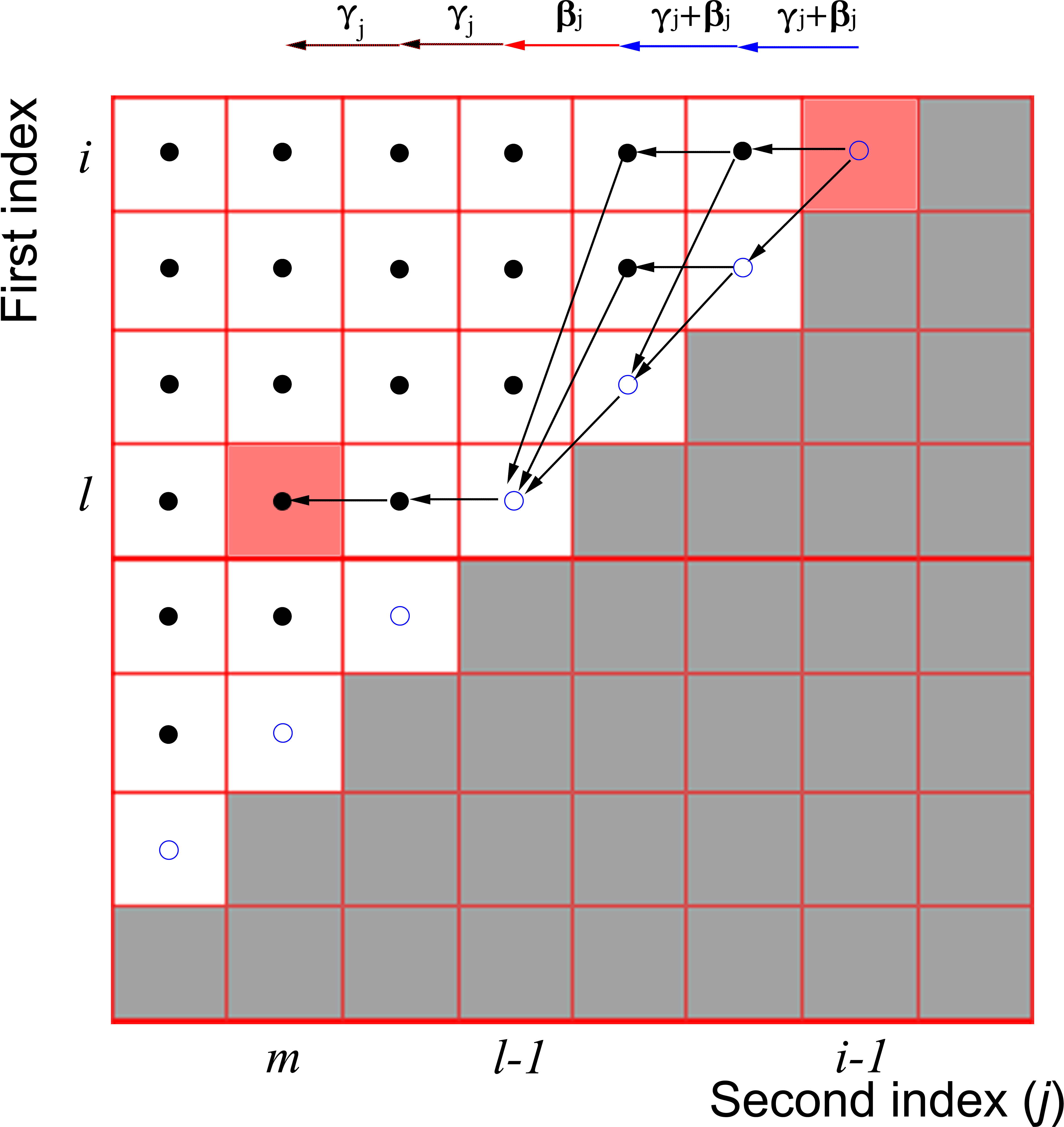}
\caption{Graph of possible trajectories to pick up $\Delta \mathbf{x}_{l,\overline{m}}^2$ like terms from a decomposition of the vector $||\boldsymbol{\eta}_i||^2 = \Delta \mathbf{x}_{i,\overline{i-1}}^2$. The dots represent possible indices of intermediate $\Delta \mathbf{x}_{k,\overline{j}}^2$ terms and the blue dots are those that correspond to separation vectors $\boldsymbol{\eta}_j$. In order to get from $\Delta \mathbf{x}_{i,\overline{i-1}}^2$ to $\Delta \mathbf{x}_{l,\overline{m}}^2$ the index trajectories must pass through $\Delta \mathbf{x}_{l,\overline{l-1}}^2$ and thus $a_{l,m}$ is the product over $j$ of three separate regions $m<j<l$, $j=l$ and $l<j<i$ unless $l=i$ and there is only one trajectory.}\label{graphdiag}
\end{center}
\end{figure}

\begin{align}
\Delta \mathbf{x}_{i,\overline{j}}^2 &=  \notag \frac{\overline{d}_{j-1}^2}{\overline{d}_j^2} \Delta \mathbf{x}_{j,\overline{j-1}}^2 + \Delta \mathbf{x}_{i,j}^2 - \frac{\overline{d}_{j-1}}{\overline{d}_j} \left(2\Delta \mathbf{x}_{i,j}\Delta \mathbf{x}_{j,\overline{j-1}} \cos(\theta_{i,j})\right) \\
\notag &= \frac{\overline{d}_{j-1}^2}{\overline{d}_j^2} \Delta \mathbf{x}_{j,\overline{j-1}}^2 + \Delta \mathbf{x}_{i,j}^2 - \frac{\overline{d}_{j-1}}{\overline{d}_j} \left(\Delta \mathbf{x}_{j,\overline{j-1}}^2 + \Delta \mathbf{x}_{i,j}^2 - \Delta \mathbf{x}_{i,\overline{j-1}}^2   \right) \\
\label{triangleeqn} &= \beta_j \Delta \mathbf{x}_{j,\overline{j-1}}^2    +   \gamma_j \Delta \mathbf{x}_{i,\overline{j-1}}^2     + \omega_j  \Delta \mathbf{x}_{i,j}^2 ,
\end{align}
where $\beta_j = d_j\overline{d}_{j-1} / \overline{d}_j^2 $, $\gamma_j = \overline{d}_{j-1} / \overline{d}_j $ and $\omega_j = d_j / \overline{d}_j$. By definition of the separation vectors $||\boldsymbol{\eta}_i||$, we have that
\begin{equation} \label{etai_iterate}
||\boldsymbol{\eta}_i||^2 = \Delta \mathbf{x}_{i,\overline{i-1}}^2 = \beta_{i-1} \Delta \mathbf{x}_{i-1,\overline{i-2}}^2    +   \gamma_{i-1} \Delta \mathbf{x}_{i,\overline{i-2}}^2     + \omega_{i-1}  \Delta \mathbf{x}_{i,i-1}^2.
\end{equation}
We apply successive iterations of Equation (\ref{triangleeqn}) to Equation (\ref{etai_iterate}) to eliminate all distances including a centre of diffusion (distances with an overbar over the second index). Since indices on the RHS of Equation (\ref{triangleeqn}) are less than or equal to the indices of the LHS, we have that
\begin{equation}\label{etaisq}
||\boldsymbol{\eta}_i||^2 = \sum_{l=2}^i\sum_{m=1}^{l-1} a_{l,m}  \Delta \mathbf{x}_{l,m}^2,
\end{equation}
where $a_{l,m}$ can be constructed by considering all terms that contain $\Delta \mathbf{x}_{l,m}^2$ after iteration of Equation (\ref{triangleeqn}). We notice that the RHS of Equation (\ref{triangleeqn}) has three terms, one that is already a distance between two molecules (of the form $\Delta \mathbf{x}_{l,m}^2$) and two others that require further iteration. With each iteration, the second index $j$ is reduced by 1 but the first index is either left unchanged or drops in line with $||\boldsymbol{\eta}_{j-1}||^2$. Each time the first index is left unchanged, a factor of $\gamma_j$ is multiplied a term in to the coefficient $a_{l,m}$ and each time the first index drops in line with $||\boldsymbol{\eta}_{j-1}||^2$, a factor of $\beta_j$ is multiplied to a term in  the coefficient $a_{l,m}$. The coefficient $a_{l,m}$ is therefore the product of all of the $\gamma_j$ and $\beta_j$ steps in going from $\Delta \mathbf{x}_{i,\overline{i-1}}^2$ to $\Delta \mathbf{x}_{l,\overline{m}}^2$ (that is, the product of $\gamma_j$ or $\beta_j$ over second index values $j$ for each reduction on the interval $m< j <i$) and then summed over all possible trajectories the indices may take and finally multiplied by $\omega_m$ to go from $\Delta \mathbf{x}_{l,\overline{m}}^2$ to $\Delta \mathbf{x}_{l,m}^2$. Figure \ref{graphdiag} illustrates the trajectories that the indices may take in going from $(i,\overline{i-1})$ to a particular $(l,\overline{m})$ where $m< j <i$. It is important to note that all trajectories must pass through the indices $(l,\overline{l-1})$ and thus must have a $\beta_l$ step and $(l-m-2)$ $\gamma_j$ steps as $j$ decreases each iteration between $j=l-1$ and $j=m+1$. Finally, it does not matter which, either $\gamma_j$ or $\beta_j$, step is taken for $l< j<i$ and thus we should multiply $\gamma_j + \beta_j$ for each iteration in this region to obtain the sum of all possible trajectories. Noting, finally, that if $l=i$ we only have $\gamma_j$ steps, 
\begin{align}\notag
a_{l,m} &= 
  \left\lbrace 
    \begin{array}{cc}
      \omega_m\beta_l\left[\prod_{j=m+1}^{l-1} \gamma_j\right]\left[\prod_{j=j+1}^{i-1} \gamma_j+\beta_j \right] & \ \text{ for } l\neq i \\
      \omega_m\left[\prod_{j=m+1}^{i-1} \gamma_j\right] & \ \text{ for } l = i
    \end{array}
\right. , \\ 
&= 
  \left\lbrace 
    \begin{array}{cc}
      -d_m d_l/\overline{d}_{i-1}^2 & \ \text{ for } l\neq i \\
      d_m/\overline{d}_{i-1} & \ \text{ for } l = i
    \end{array}
\right. .
\end{align}
Substituting $a_{l,m}$ into Equation (\ref{etaisq}) gives 
\begin{equation}\label{whatiseta}
||\boldsymbol{\eta}_i||^2 = \sum_{m=1}^{i-1} \frac{d_m}{\overline{d}_{i-1}}  \Delta \mathbf{x}_{i,m}^2   - \sum_{l=2}^{i-1}\sum_{m=1}^{l-1} \frac{d_m d_l}{\overline{d}_{i-1}^2}  \Delta \mathbf{x}_{l,m}^2.
\end{equation}
Substituting Equation (\ref{whatiseta}) into Equation (\ref{appendeqn}) gives
\begin{equation}
\frac{\mathcal{P}_N^2}{\Delta_N} = \sum_{i=2}^N \sum_{m=1}^{i-1} \frac{d_m d_i}{\overline{d}_{i}}  \Delta \mathbf{x}_{i,m}^2   - \sum_{i=2}^N \sum_{l=2}^{i-1}\sum_{m=1}^{l-1} \frac{d_m d_l d_i}{\overline{d}_{i}\overline{d}_{i-1}}  \Delta \mathbf{x}_{l,m}^2.
\end{equation}
Interchanging the sum over $i$ and the sum over $l$ in the second term on the RHS and noting that 
$$
\sum_{i=l+1}^N \frac{d_i}{\overline{d}_i\overline{d}_{i-1}} = \frac{\overline{d}_N - \overline{d}_l}{\overline{d}_N\overline{d}_l},
$$
which can be proven easily by induction for all $N>l$, we have
\begin{equation}
\frac{\mathcal{P}_N^2}{\Delta_N} = \sum_{i=2}^N \sum_{m=1}^{i-1} \frac{d_m d_i}{\overline{d}_{i}}  \Delta \mathbf{x}_{i,m}^2   -  \sum_{l=2}^{N-1}\sum_{m=1}^{l-1} \frac{d_m d_l (\overline{d}_N - \overline{d}_l) }{\overline{d}_N \overline{d}_l}  \Delta \mathbf{x}_{l,m}^2.
\end{equation}
Separating the $N$-th term in the first sum of the first term, relabelling the dummy index $l$ with $i$ in the second term and joining the remaining double summations together gives
\begin{align}
\frac{\mathcal{P}_N^2}{\Delta_N} &=  \sum_{m=1}^{N-1} \frac{d_m d_N}{\overline{d}_{N}}  \Delta \mathbf{x}_{N,m}^2  + \sum_{i=2}^{N-1}\sum_{m=1}^{i-1} \frac{d_m d_i  }{\overline{d}_N }  \Delta \mathbf{x}_{i,m}^2, \\
&=   \sum_{i=2}^{N}\sum_{m=1}^{i-1} \frac{d_m d_i  }{\overline{d}_N }  \Delta \mathbf{x}_{i,m}^2 \\
&=   \sum_{i>m} \frac{d_m d_i  }{\overline{d}_N }  \Delta \mathbf{x}_{i,m}^2,
\end{align}
where the summation in the final statement is for all combinations of $i$ and $m$ such that $i>m$ and all molecule combinations are only counted once. This expression for $\mathcal{P}_N$ is clearly independent on the order in which the molecules are labelled. The scale factor $$\Delta_N =   \frac{\overline{d}_N }{\sum_{i>m} d_m d_i  },$$ as presented in Equation (\ref{scalefact}), was chosen arbitrarily so that the resultant square proximity $\mathcal{P}_N^2$ would be a weighted average of square molecular separations (which best fits the intuitive physical definition of this metric) and consequently, for $N=2$, $\mathcal{P}_2^2 = \Delta \mathbf{x}_{2,1}^2$ is consistent with the classical Smoluchowski theory.

\end{document}